\documentclass[preprint,aps,showpacs,amsmath,nofootinbib,eqsecnum]{revtex4}
\usepackage{epsfig}
\usepackage{amsmath}
\usepackage{amsfonts}
\usepackage{amssymb}
\usepackage{color}

\def\bc{\begin{center}}
\def\nno{\nonumber}
\def\ec{\end{center}}
\def\be{\begin{eqnarray}}
\def\ee{\end{eqnarray}}

\newcommand{\omits}[1]{}

\newcommand{\vect}[1]{\mbox{\boldmath{$ #1$}}}

\definecolor{dyellow}{rgb}{1.,0.8,.0}
\definecolor{myblue}{rgb}{.1,.1,.7}
\definecolor{dcyan}{rgb}{.0,.6,.6}
\definecolor{dmagenta}{rgb}{0.6,0.0,0.6}
\definecolor{brown}{rgb}{0.6,0.2,0.}
\definecolor{darkblue}{rgb}{.0,.0,0.5}
\definecolor{darkred}{rgb}{0.75,0.0,0.0}
\definecolor{orange}{rgb}{1.,.6,.0}
\definecolor{dorange}{rgb}{0.8,.4,.0}
\definecolor{lightgray}{rgb}{0.7,0.7,0.7}
\definecolor{darkgreen}{rgb}{0.0,0.6,0.0}
\definecolor{purple}{rgb}{.4,.0,.4}
\def\Dl{\Delta}
\def\La{\Lambda}

\def\al{\alpha}

\def\dl{\delta}
\def\eps{\epsilon}
\def\ka{\kappa}
\def\la{\lambda}

\def\si{\sigma}

\def\d#1#2{\frac{\displaystyle #1}{\displaystyle #2}}
\def\r{\partial}

\newcommand{\dS}{$d{S}$}
\newcommand{\Mink}{${M}ink$}
\newcommand{\AdS}{${A}dS$}
\newcommand{\BdS}{${B}d{S}$}

\newcommand{\HlsM}{${H}_l \subset {M}^{1,4}$}
\newcommand{\HrsM}{${H}_R \subset {M}^{1,4}$}

\newcommand{\GR}{general relativity}% {${\cal GR}$}
\newcommand{\SR}{special relativity}%{${\cal SR}$}
\newcommand{\dSR} {$d{S}$ special relativity}

\newcommand{\HsM}{${H}_l \subset M^{1,4}$}

\newcommand\btd{\raise 2pt
\hbox{$\hat\bigtriangledown$}\hskip 1.5pt}
\newcommand\bt{\raise 2pt
\hbox{$\bigtriangledown$}\hskip 1.5pt}

\newcommand{\FLT}{$FLT$}

\def\PRD{{\it Phys. Rev.}~{\bf D}}
\def\PRL{{\it Phys. Rev. Lett }}
\def\PLA{{\it Phys. Lett.}~{\bf A}}

\def\CTP{{\it Commun. Theor. Phys. }}

\newenvironment{proof}{\textbf{Proof:}}{\rule{2mm}{4mm}}

\begin{document}

%\begin{frontmatter}
\title{Snyder's Model -- de Sitter Special Relativity Duality\\
and de Sitter Gravity}%\vskip 3mm %[10mm]

\author{{Han-Ying Guo}$^{1,2}$}
\email{hyguo@itp.ac.cn}
\author{{Chao-Guang Huang}$^{1}$}
\email{huangcg@mail.ihep.ac.cn}
\author{{Yu Tian}$^{3}$}
\email{ytian@bit.edu.cn}
\author{{Hong-Tu Wu}$^{4}$}
\email{lobby_wu@yahoo.com.cn}
\author{{Zhan Xu}$^{5}$} \email{zx-dmp@tsinghua.edu.cn}
\author{{Bin Zhou}$^{6}$} \email{zhoub@bnu.edu.cn}

%\affiliation{%
%${}^1$ CCAST (World Laboratory), P.O. Box 8730, Beijing 100080,
%China,}
\affiliation{%
${}^1$ Institute of High Energy Physics, Chinese Academy of
Sciences, P.O. Box 918-4, Beijing 100049, China,}
\affiliation{%
${}^2$ Institute of Theoretical Physics, Chinese Academy of
Sciences, Beijing 100080, China,}
\affiliation{%
${}^3$ Department of Physics, Beijing Institute of Technology,
Beijing 100081, China,}
\affiliation{%
${}^4$Department of Mathematics, Capital Normal University, Beijing
100037, China,}
\affiliation{%
${}^5$Physics Department, Tsinghua University, Beijing 100084,
China}
\affiliation{%
${}^6$Department of Physics, Beijing Normal University, Beijing
100875, China.}

\date{June, 2007}

\begin{abstract}
Between Snyder's quantized space-time model in de Sitter space of
momenta and the \dS\ special relativity on \dS-spacetime of radius
$R$ with Beltrami coordinates, there is a one-to-one dual
correspondence supported by a minimum uncertainty-like argument.
Together with Planck length $\ell_P$, $R\simeq (3/\Lambda)^{1/2}$
should be a fundamental constant. They lead to a dimensionless
constant $g{\sim\ell_PR^{-1}}=(G\hbar c^{-3}\Lambda/3)^{1/2}\sim
10^{-61}$. These indicate that physics at these two scales should be
dual to each other and there is in-between gravity of local
\dS-invariance characterized by $g$. A simple model of \dS-gravity
with a gauge-like action on umbilical manifolds may show these
characters. It can pass the observation tests and support the
duality.
\end{abstract}


 %Euclidean and projective geometry

%(Please choose 4 pacs numbers from the above 6. The LaTeX code gives
%the explanation of the pacs numbers.)
%}
%\keywords{Planck scale-cosmological constant duality, de Sitter invariant special relativity, Snyder model, de Sitter gravity}% keywords here, in the form: keyword \sep keyword

%\end{frontmatter}

\maketitle
\tableofcontents

%\newpage

\section{Introduction}

Long time ago, Snyder \cite{Snyder} proposed a quantized space-time
model. In his model, Snyder started with a projective geometry
approach to the de Sitter(\dS)-space of momenta with a scale $\al$,
which is proportional to $\hbar/a$ in \cite{Snyder}, near or equal
to the Planck {scale}. The energy and momentum of a particle were
identified with the inhomogeneous projective coordinates. Then, the
spacetime `coordinates' were naturally defined as 4-`translation'
generators $\hat x_\mu$ of \dS-algebra $\mathfrak{so}(1,4)$ in
\dS-space of momenta and became noncommutative. It is important that
there is a new kind of uniform `velocity' motions with constant
`group velocities' for some `wave packets' in the model. In
addition, it may indicate that correspondingly there is a new kind
of uniform coordinate velocity motions for some particles in the
\dS-spacetime
\cite{Lu,LZG,BdS,BdS05,IWR,T,NH,Lu05,PRdual,PoI,C3,Yan}. But, this
was not recognized.

Recently, in order to explain the Greisen-Zatsepin-Kuz'min effects
the doubly/deformed special relativity (DSR) has been proposed
\cite{DSR}. In DSR, there is also a large energy-momentum scale
$\kappa$ near the Planck energy, related to $a$ in Snyder's model in
addition to $c$. Soon, it is found that there is a close relation
between Snyder's model and DSR. In fact, DSR can be regarded as
generalization of Snyder's model \cite{DSRdS}. And, most DSR models
with $\kappa$-Poincar\'e algebra can be realized geometrically by
means of particular coordinate systems on
\dS/Minkowski(\Mink)/\AdS-space of momenta \cite{DSRdS} other than
the inhomogeneous projective coordinates used by Snyder's model in
\dS-space of momenta. Thus, there is a kind of coordinate
transformations from Snyder's model to some of DSR on \dS-space of
momenta and vise versa.

In fact, the projective geometry  approach to \dS-space is basically
equivalent to the Beltrami-like model (Beltrami model for short).
Historically, de Sitter \cite{dS17} first used the Beltrami
coordinates \cite{beltrami,R} for his  solution of constant
curvature, the \dS-spacetime, in the course of debate with Einstein
on `relative inertia'. Later, Pauli \cite{Pauli20} mentioned this
metric in Euclidean signature.

In his first paper \cite{1905}, Einstein assumed the rigid ruler at
rest be Euclidean. For  free space in large scale, it has  less
observation basis since it is not supported by the asymptotic
behavior of  our universe \cite{Riess98,WMAP}. Actually, once
Einstein's Euclid assumption is released, there should be three
kinds of \SR\ with $ISO(1,3), SO(1,4), SO(2,3)$-invariance on
\Mink/\dS/\AdS-spacetime, respectively. This is in analogy with the
remarkable historic issue on Euclidean fifth axiom. Once the axiom
is weakened, there are Euclid, Riemann and Lobachevski geometries of
zero, positive and negative constant curvature  on an almost equal
footing. In these 4-dimensional geometries, say, there is a kind of
special coordinate systems, respectively.  In these coordinates, the
points, straight-lines of linear form, metric and other geometric
objects are invariant or transformed among themselves under  linear
transformations of $ISO(4)$-invariance or fractional linear
transformations with a common denominator (\FLT s) of  $SO(5),
SO(1,4)$-invariance, respectively. For Lobachevski plane, Beltrami
\cite{beltrami} first introduced such coordinates. Then, changing
the metric to physical signature by an inverse Wick rotation
\cite{IWR}, these spaces become \Mink-, \dS-\ and \AdS-spacetime
with invariance of $ISO(1,3), SO(1,4)$ and $SO(2,3)$, respectively.
And the geometric objects such as points, straight-lines and  Euclid
or Beltrami-metric become corresponding events, straight world-lines
and \Mink- or Beltrami-metric with signature of $(+,-,-,-)$,
respectively.

For Euclid's counterpart, there is  Einstein's special relativity in
\Mink-space based on the principle of relativity and the postulate
on universal invariant of the speed of light $c$. What should
correspond to the other two non-Euclidean counterparts? Those are
just two other kinds of \SR\ on \dS/\AdS-space based on the
corresponding principle of relativity and the postulate on universal
constants of the speed of light $c$ and the curvature radius $R$.

More concretely, say, for \dS-space with radius $R$, Beltrami
coordinates are in analogy with \Mink-coordinates on \Mink-space. It
is precisely the Beltrami model of a \dS-hyperboloid ${H}_R$ in a
5-d \Mink-space \dS\ $\backsimeq {H}_R \subset M^{1,4}$. Via
Beltrami coordinate atlas, the \dS-space can be covered patch by
patch, in which particles and light signals move along the timelike
or null geodesics being straight world-lines with {\it constant}
coordinate velocities in each patch, respectively. And all these
properties are invariant under \FLT s of $SO(1,4)$ symmetry among
Beltrami-systems. In the light of inertial motions in both Newtonian
mechanics and Einstein's special relativity, these particles and
signals should be in free motion of inertia without gravity.
Accordingly, the Beltrami coordinates and observers should also be
of inertia and there should be the principle of relativity in
\dS-spacetime.

In 1970, Lu \cite{Lu} first noticed these important properties and
began to study \dS/\AdS-invariant special relativity on
\dS/\AdS-space (\dS/\AdS\, special relativity for short),
respectively \cite{LZG}. Recently, promoted by the observations on
our dark universe, the studies are being made further
\cite{BdS,BdS05,IWR, T, NH, Lu05,PRdual,PoI,C3,
Yan}\footnote{Recently, another version of \dS\ special relativity
has been proposed \cite{AAP}, but the principle of relativity,
inertial frames and the transformations among them are not mentioned
there.}.

It is interesting to see that in terms of the Beltrami model of
\dS-space (denoted as \BdS-space)
\cite{dS17,Pauli20,IWR,Lu,LZG,BdS,BdS05}, there is a dual one-to-one
correspondence  between Snyder's quantized space-time model
\cite{Snyder} as a DSR \cite{DSR,DSRdS} and the \dS\, special
relativity \cite{IWR,Lu,LZG,BdS,BdS05,T}. {Actually, the \dS\
special relativity can be  regarded and simply formulated as a
spacetime-counterpart of Snyder's model for \dS-space of momenta so
long as the constant $\al$ in Snyder's model as \dS-radius of
momenta near the Planck scale is replaced by $R$ as radius of
\dS-spacetime.} Furthermore, via the constant $a$ in Snyder's model
(or $\kappa$ in DSR) or Planck length $\ell_P$ and the cosmological
constant $\Lambda$, a dimensionless constant $g$ can be introduced:
\be \label{g}%
\iota^{2}:=\kappa^2/R^2 \to {g^2 \sim
G\hbar c^{-3}\Lambda /3 = {\ell_P}^2R^{-2}} \simeq 10^{- 122}.
\ee%
Since Newton's gravitational constant appears, $g$ should describe
gravity with local \dS-invariance between these two scales. In fact,
this dimensionless constant $g$ has been appeared in a simple model
of \dS-gravity with \dS-algebra as gauge algebra
\cite{dSG,uml,T77,QG} to characterize self-interaction of gravity
with local \dS-invariance. In addition, an
uncertainty-like argument can be given via a `tachyon' dynamics in
embedded space, which may support the dual one-to-one
correspondence.

Based upon these important properties we may expect that there
should be a duality between {physics at the scale of} Planck length
$\ell_p=(G\hbar c^{-3})^{1/2}$ and \dS-radius $R=(3/\Lambda)^{1/2}$
with the cosmological constant $\Lambda$ {regarded as} a fundamental
one in the nature. {That is,} physics at such two fundamental scales
are dual to each other in some `phase' space, and in-between there
is gravity of local \dS-invariance characterized by the
dimensionless constant $g$.

Thus, there no longer exist the puzzle on $\La$ as {the ordinary
`vacuum energy' in the viewpoint of \dS\ \SR\ and the duality above.
This is due to that the concept of `vacuum' now is {\dS\ invariant}
and that the so-called `vacuum energy' {calculated} in the
\Mink-space becomes improper. In the viewpoint of \dSR, the puzzle
should be: w}hat is the origin of the dimensionless constant $g$
{and i}s it calculable?

In the point of view of \GR, there is no room for the \SR\ on
\dS/\AdS-space. In the point of view of \dS/\AdS\, \SR, however,
there is no gravity on \dS/\AdS-space. As is just explained, this
`funny' stuff in Einstein's relativity is in analogy with the
remarkable historic issue on Euclid fifth axiom. Thus, we should
explain how to describe  gravity in {the point of} view of
\dS/\AdS\, \SR.

In parallel with the local Poincar\'e gauge theory of gravity
\cite{EC,held,An}, we suggest that gravity should be based on the
localization of \dS\, special relativity and described by a
gauge-like dynamics characterized by the dimensionless constant $g$.
We show how to localize the \dS-hyperboloid \HrsM\ at each events on
a kind of umbilical manifolds of local \dS-invariance and also very
briefly introduce a simple model with a Yang-Mills type action
\cite{dSG,uml,T77,QG} of such gravity with local \dS-invariance. We
show that this model support the duality and also provide some hints
on above questions on the dimensionless constant $g$.

This paper is arranged as follows. We first review the general
properties of the Beltrami model of 4-d Riemann-sphere and of the
Beltrami model of \dS-space via an inverse Wick rotation in Sec. II.
Next, we recall some important relevant issues in \dS\, special
relativity in terms of the Beltrami model of \dS-spacetime and in
Snyder's model of quantum space-time (together with DSR) in terms of
the Beltrami model of \dS-space of momenta, respectively, in Sec.
III and IV. Then, we show that there is a dual one-to-one
correspondence between Snyder's model and  \dS\, special relativity,
which is supported by a minimum uncertainty-like argument indicated
by a `tachyon' dynamics and propose the duality for physics at and
in-between the two scales in Sec. V. In Sec. VI we explain how to
describe gravity based on localization of \dS\, \SR\ and introduce a
simple model of \dS-gravity briefly on a kind of umbilical
manifolds. Finally, we end with some concluding remarks.

%%%%%%%%%%%%%%%%%%%%%%%%%%%%%%%%%%%%%%%%%%%%%%%%%
%
%              Beltrami-model
%
%%%%%%%%%%%%%%%%%%%%%%%%%%%%%%%%%%%%%%%%%%%%%%%

\section{Beltrami Model of Riemann-Sphere and de Sitter Spacetime}

\subsection{Beltrami Model of Riemann-Sphere}

A 4-d Riemann-sphere ${ S}^4$ can be embedded in a 5-d Euclid space
$E^5$
\be\label{4s}%
{ S}^4:&&\delta_{AB}\xi^A \xi^B=l^2>0, \quad A, B=0, \cdots, 4,\\\label{5ds}%
&& ds_E^2=\delta_{AB}d\xi^A d\xi^B=d\xi ^t{\cal I}d\xi. %
\ee%
 They are invariant under rotations of $SO(5)$:
\be\label{so5}%
 \xi~ \rightarrow ~\xi'=S~\xi, \quad S^t {\cal I} S={\cal
I}, ~~\forall ~S ~\in ~{SO(5)}.
\ee%

A Beltrami model ${ B}_l$ of ${ S}^4$ is the intrinsic geometry of
${ S}^4$ with Beltrami coordinate atlas.
In a patch, say  $U_{+4}$, %
\be\label{Bcrd}%
x^\mu:=l\frac{\xi^\mu}{\xi^4}, \quad \xi^4{>} 0 ,\quad
\mu =0,\cdots, 3.%
\ee%
To cover  ${ B}_l\sim$ ${ S}^4$, one patch is not enough, but all
properties of ${ S}^4$ are well-defined in the ${ B}_l$ patch by
patch with
\be\label{sigma}%
\sigma_E(x):=\sigma_E(x,x)=1+l^{-2}\delta_{\mu\nu}x^\mu x^\nu>0,\qquad\\\label{4Bds}
ds_E^2=\{\delta_{\mu\nu}\sigma_E^{-1}(x)-l^{-2}\sigma_E^{-2}(x)\delta_{\mu \la}x^\la\delta_{\nu\ka}x^\ka\}
dx^\mu dx^\nu.
\ee%

It is clear that the inequality (\ref{sigma}) and the Beltrami
metric (\ref{4Bds}) are invariant under \FLT s among Beltrami
coordinates $x^\mu$, which can be written in a transitive
form  sending the point $A(a^\mu)$ to the origin $O(o^\mu=0)$,%
\be\label{FLT} x^\mu\rightarrow
\tilde{x}^\mu&=&\pm\sigma_E^{1/2}(a)\sigma_E^{-1}(a,x)(x^\nu-a^\nu)N_\nu^\mu,\\\nno
N_\nu^\mu&=&O_\nu^\mu-l^{-2}%
\delta_{\nu\ka}a^\ka a^\la
(\sigma_E(a)+\sigma_E^{1/2}(a))^{-1}O_\la^\mu,\\\nno
O&:=&(O_\nu^\mu)\in SO(4).%
\ee

There is an invariant for two points $A(a^\mu)$ and $B(b^\mu)$  on
${ B}_l$ %
\be\label{AB} %
{\Delta}_{E,l}^2(a, {b}) = -l^2
[\sigma_E^{-1}(a)\sigma_E^{-1}({b})\sigma_E^2(a,{b})-1].%
\ee %
The arc-length of the geodesic segment $\overline{AB}$ connecting
$A$ and $B$ is
\be \label{ABL}%
L(A,B)= \int_A^B ds_E =l \arcsin (|\Delta_{E{,l}}(a,b)|/l).%
\ee
It may also be written as
\be\label{action}%
 L(A,B)=\int_A^B ds_E=\int_{a^0}^{b^0} dx^0
\sqrt{g_{\mu\nu}\dot{x}^\mu\dot{x}^\nu }, \ee
where $\dot{x}^\mu:=dx^\mu/dx^0$ and $g_{\mu\nu}$ is the
Beltrami-metric.  From its variation, it follows a geodesic
equation or alternatively
\be
\delta L(A,B)&=&\int_{a^0}^{b^0}dx^0 \{R^2{\cal R}_{i \ka \la
\mu}\frac{dx^\ka}{ds}\frac{dx^\la}{ds}\frac{d}{ds}(\frac{dx^\mu}{dx^0})\delta
x^i\nno \\
&&+\frac{d}{dx^0}(\frac{dx^\mu}{ds}g^{\mu i}\delta x^i)
\} , \quad i=1,2,3,  %
\ee
where ${\cal R}_{\nu \ka \la \mu}$ is  Riemann curvature tensor.
$\dl L(A,B) =0$ and $\dl x^i=0$ on the initial and final
hypersurfaces give rise to the equation of motion
\be%
\frac{d}{ds}\frac{dx^i}{dx^0}=0,%
\ee
which results in
\begin{eqnarray}\label{um}%
\frac{dx^i}{dx^0}=consts.%
\end{eqnarray}%
Integrating it further gives
\be\label{sl}%
x^i=\al^i x^0+\beta^i;\quad \al^i,\beta^i=consts.%
\ee%
Namely, the geodesics as shortest curves in  Beltrami model of
Riemann sphere are straight-lines in linear form. In fact, Eq.
(\ref{um}) can be
obtained directly from the first integral of geodesic equation,%
\be\label{q}%
{q^\mu}:=\sigma_E^{-1}(x)\frac{dx^\mu}{ds}=consts.%
\ee%

It is important that all these results are invariant under the
\FLT s (\ref{FLT}) of $SO(5)$ and globally true in Beltrami
atlas patch by patch.

In  terminology of projective geometry, Beltrami coordinates are
inhomogeneous projective ones. But,  antipodal identification should
not be taken here in order to preserve orientation. The great
circles on 4-d Riemann{-}sphere (\ref{4s}) are mapped to
straight-lines (\ref{sl}) in ${ B}_l$, and vice versa.

\subsection{The Beltrami model of \dS-spacetime via an inverse Wick rotation}

From an inverse Wick rotation \cite{BdS,IWR} of {the 5-d embedded
space, which turns $\xi^0$ to be time-like, the Riemann-sphere $S^4
\subset E^5$ (\ref{4s}) becomes} the \dS-hyperboloid \HlsM:
\be\label{qiu2}%
{H}_l&:&\eta_{AB}\xi^{A}\xi^{B}=\xi^t {\cal J}\xi=-l^2,\\
&&ds^2=\eta_{A B} d\xi^A d\xi^B=d\xi^t {\cal J} d\xi, \label{metric-qiu2}\\
\partial_P H_l&:&\eta_{AB}\xi^{A}\xi^{B}=\xi^t {\cal J}\xi=0, \  A,B=0,\cdots,4,%
\label{bound-qiu2}\ee%\
where ${\cal J}=(\eta_{A B})={\rm diag}(1,-1,-1,-1,-1 )$ and
$\partial_P$ denotes the projective boundary. They are invariant
under \dS-group $SO(1,4)$:%
 \be\label{dst}%
 \xi~ \rightarrow ~\xi'=S~\xi, ~~
S^t {\cal J} S={\cal J},~~ \forall ~S ~\in ~{SO(1,4)}.
\ee%

The great circles on the Riemann-sphere (\ref{4s}) now transfer to
a kind of uniform `great circular' motions of a
particle with `mass' $m_l$ characterized by a conserved 5-d
angular momentum on \HlsM: %
\be\label{angular5a}%
 \frac{d{\cal
L}^{AB}}{ds}=0,\quad {\cal
L}^{AB}:=m_{l}(\xi^A\frac{d\xi^B}{ds}-\xi^B\frac{d\xi^A}{ds}),%
\ee%
with an Einstein-like formula for `mass' $m_{l}$
\begin{eqnarray}\label{emla}%
-\frac{1}{2l^2}{\cal L}^{AB}{\cal L}_{AB}=m_{l}^2,\quad%
{\cal L}_{AB}=\eta_{AC}\eta_{BD}{\cal L}^{CD}.
\end{eqnarray}

Further, a `simultaneous' 3-hypersurface of $\xi^0={\rm const}$ is an expanding $S^3$:%
\be\label{s3}%
\delta_{IJ}\xi^I\xi^J&=&l^2+(\xi^0)^2,~~ I, J=1,\cdots, 4;\\\nonumber%
dl^2&=&\delta_{IJ}d\xi^I d\xi^J.%
\ee%

The generators of  $so(5)$-algebra become the ones of \dS-algebra
$so(1,4)$, which read ($\hbar=1$)
\be\label{Generator}%
{\hat{\cal L}}_{AB} = \frac 1 i\left(\xi_A
\frac{\partial}{\partial\xi^B} - \xi_B
\frac{\partial}{\partial\xi^A}\right),\quad \xi_A=\eta_{AB}\xi^B,
\ee%
or the Killing vector fields (without $i$) on \dS-hyperboloid. They
are globally defined on the \dS-hyperboloid.

The first Casimir operator of the algebra is%
\be\label{C15}%
\hat C_1:= -\frac 1 2 l^{-2} \mathbf{\hat {\cal L}}_{AB}
\mathbf{\hat {\cal L}}^{AB},\quad
\mathbf{\hat {\cal L}}^{AB}:=\eta^{AC}\eta^{BD} \mathbf{\hat {\cal L}}_{CD},%
\ee%
with eigenvalue $m^2_l$, which gives rise to the classification of
the `mass' $m_l$.

It is clear that the \BdS-space can be given  either by the
Beltrami model of Riemann-sphere via an inverse Wick rotation or
by the generalized `gnomonic' projection without antipodal
identification of the \dS-hyperboloid \HlsM.

Thus, there exists Beltrami coordinate atlas covering \dS-space
patch by patch. On each patch, there are condition and Beltrami
metric with $\eta_{\mu \nu}={\rm diag}(1,-1,-1,-1)$:
\be\label{domain}%
\sigma(x)=\sigma(x,x):=1-l^{-2} \eta_{\mu\nu}x^\mu x^\nu
>0,\qquad\\\label{metric} %
ds^2=[\eta_{\mu\nu}\sigma^{-1}(x)+ l^{-2}
\eta_{\mu\la}\eta_{\nu\ka}x^\la x^\ka
\sigma^{-2}(x)]dx^\mu dx^\nu. %
\ee%
Under $FLT$s of $SO(1,4)$,
\be\label{G}\nno%
x^\mu\rightarrow \tilde{x}^\mu&=&\pm
\sigma^{1/2}(a)\sigma^{-1}(a,x)(x^\nu-a^\nu)D_\nu^\mu,\\
D_\nu^\mu&=&L_\nu^\mu+ l^{-2}%
\eta_{\nu \la}a^\la a^\ka
(\sigma(a)+\sigma^{1/2}(a))^{-1}L_\ka^\mu,\\\nno
L&:=&(L_\nu^\mu)\in SO(1,3),%
\ee
which transform a point $A(a^{\mu})$ with $\sigma(a^{\mu})>0$ to
the origin, the system $S(x)$ transforms to $\tilde S(\tilde x)$ and
the inequality (\ref{domain}) and Eq.(\ref{metric}) are invariant.

In such a \BdS, the generators of $FLT$s or the Killing vectors read
\be\label{generator}%
  {\hat q}_\mu =(\delta_\mu^\nu-l^{-2}x_\mu x^\nu) \partial_\nu,~~
  x_\mu:=\eta_{\mu\nu}x^\nu,\qquad\\
  {\hat L}_{\mu\nu} = x_\mu {\hat q}_\nu - x_\nu {\hat q}_\mu
  = x_\mu \partial_\nu - x_\nu \partial_\mu \in so(1,3), \label{Lorentz}
\ee%
and form an $so(1,4)$ algebra
\be\label{so14} \nno%
  [ \hat{q}_\mu, \hat{q}_\nu ] = l^{-2} \hat{L}_{\mu\nu},~~
  {[} \hat{L}_{\mu\nu},\hat{q}_\ka {]} =
    \eta_{\nu\ka} \hat{q}_\mu - \eta_{\mu\ka} \hat{q}_\nu,\\
  {[} \hat{L}_{\mu\nu},\hat{L}_{\ka\la} {]} =
    \eta_{\nu\ka} \hat{L}_{\mu\la}
  - \eta_{\nu\la} \hat{L}_{\mu\ka}
  + \eta_{\mu\la} \hat{L}_{\nu\ka}
  - \eta_{\mu\ka} \hat{L}_{\nu\la}.
\ee%

Thus, for a set of `circular observers', a set of observables that
consist of a 5-d angular momentum conserve for the uniform `great
circular' motions, satisfying an Einstein-like formula (\ref{emla})
corresponding to (\ref{C15}), etc.

{It should be mentioned that for the \dS-hyperboloid \HsM, the above
\BdS-space with Beltrami coordinates defined by Eq.(\ref{Bcrd}) is
with respect to a time-like `gnomonic' projection, i.e. $x^0$ is a
temporal coordinate, and {that} there might be other kinds of
Beltrami-like coordinate systems with respect to null and space-like
`gnomonic' projection, respectively {(see Appendix A)}. However,
only the time-like `gnomonic' projection {discussed here} should be
taken, if we require that the transformations of $SO(1,3)$, as a
subgroup of the \dS\ group, acting on the coordinates take the same
form as that of homogeneous Lorentz group in the Minkowski case or
that under $R\to \infty$ all \dS-transformations be back to the
Poincar\'e ones.}

It should also be noted that the 4-d Riemann-sphere ${S}^4$ may be
regarded as an instanton with an Euler number $e=2$ in the sense
that it is a solution of the Euclidean version of gravitational
field equations, it provides a tunneling scenario of \BdS\
\cite{IWR}. It will be shown that this is the case in a simple model
of \dS-gravity \cite{dSG,QG} as in the \GR.

%%%%%%%%%%%%%%%%%%%%%%%%%%%%%%%%%%%%%%%%%%%%%%
%
%            dS-SR
%
%%%%%%%%%%%%%%%%%%%%%%%%%%%%%%%%%%%%%%%%%%%%%%

\section{De Sitter  Special Relativity}

It has been shown that the \dS\, special relativity can be
formulated based on the principle of relativity
\cite{Lu,LZG,BdS,IWR} and postulate on universal invariants $c$
and $R$ \cite{BdS,IWR, BdS05}. In fact, the most important
properties in \dS\, special relativity can be given in a
\BdS-model with $l=R$ \cite{BdS,IWR,BdS05}.

\subsection{The law of inertia and the generalized Einstein's formula}

Why there is a law of inertia in the \dS-spacetime? This can also be
seen from another angle: the most general transformations among
inertial systems, in which a free particle moves {\it of inertia}
with constant coordinate velocities.

In fact, Umow, Weyl and Fock studied what are the most general
transformations between two inertial systems (see, e.g.
\cite{Fock}). If in an inertial system $S(x)$, an inertial motion is
described by%
\be\label{uvm}%
 x^i=x_0^i+v^i(t-t_0),\quad
v^i=\frac{dx^i}{dt}=consts,%
\ee%
and in another inertial system $S'(x')$, it may be described by%
\be\label{uvm'}%
 {x'}^i={x'}_0^i+{v'}^i(t'-t'_0), \quad
 {v'}^i=\frac{d{x'}^i}{dt'}=consts,
\ee%
Fock proved \cite{Fock} that the most general transformations
between two systems, ${x'}^\mu=f^\mu(t, x^i)$, are fractional linear
with a common denominator, which is the same as the \FLT s in
(\ref{G}). In Appendix {\ref{sect:FockThm}}, we present a new proof
for this theorem.

Thus, there is a law of inertia in Beltrami coordinates (\ref{Bcrd})
of \dS\ with curvature radius $l=R$: {\it The free particles and
light signals without undergoing any unbalanced forces should keep
their uniform motions along straight world-lines in the linear forms
in \dS-space.}

For such a free particle, there are a set of conserved observables
along a time-like geodesic:
\be\label{momt4}%
p^\mu=\sigma^{-1}(x)m_{R}^{}\frac{d x^\mu}{ds}, \quad\frac{dp^\mu}{ds}=0;
\\\label{angular4}%
 L^{\mu\nu}=x^\mu p^\nu-x^\nu p^\mu, \quad\frac{dL^{\mu\nu}}{ds}=0.%
\ee%
They are just the inverse Wick rotation counterparts of (\ref{q})
as the pseudo 4-momentum, pseudo 4-angular-momentum of the
particle and constitute a conserved 5-d angular momentum
(\ref{angular5a}).

In terms of $p^\mu$ and $L^{\mu\nu}$, the Einstein-like  formula
(\ref{emla}) becomes:
\be\label{eml}%
 -\frac{1}{2R^2}{\cal L}^{AB}{\cal
L}_{AB}=E^2-p^2{c^2}- \frac{ c^2}{2R^2} L_{(1,3)}^2=m_{R}^2{ c^4},\\
E^2=m_{R}^2c^4+{p}^2c^2 + \d {c^2} {R^2} j^2 - \d {c^4}{R^2}
k^2,\qquad%
\ee%
with energy $E$, momentum $p^i$, $p_i=\delta_{ij}p^j$, `boost'
$k^i$, $k_i=\delta_{ij}k^j$ and 3-angular momentum $j^i$,
$j_i=\delta_{ij} j^j$.  It can be proved that they are
Noether's charges with respect to the Killing vectors
(\ref{generator}).

It should be emphasized that since the generators in
(\ref{Generator}) are globally defined on the \dS-hyperboloid,
they should also be globally defined in the Beltrami atlas patch
by patch. Thus, there is a set of globally defined ten Killing
vectors in the Beltrami atlas, and correspondingly there is a set
of ten Noether's charges forming a 5-d angular momentum ${\cal
L}^{AB}$ in (\ref{angular5a}) {\it globally} in the Beltrami
atlas, though the physical meaning of each Noether's charge
depends on the Beltrami coordinate patch used. This issue will be
explored in detail elsewhere.

Note that $m^2_{R}$ now is the eigenvalue of the first Casimir
operator of \dS-group, the same as the one in (\ref{C15}) with
$l=R$. And also note that we can introduce the Newton-Hooke
constant $\nu$ \cite{NH} and link the curvature radius $R$ with
the cosmological constant $\La$
\be\label{NHc}%
\nu:=\d {c} {R} {=} \left (\d {3c^2} \Lambda \right)^{1/2},\quad
\nu^2 \sim 10^{-35}s^{-2}.
\ee%
It is very tiny {if $\Lambda$ is taken as the present value of the
dark energy\footnote{The dark energy may have other (dynamical)
contributions in addition to the cosmological constant. The
assumption that the `observed' dark energy be cosmological constant
is just a working hypothesis.}. Thus, all local experiments, whose
characteristic spatial and temporal scales are much smaller than $R$
and $\nu^{-1}$, respectively, can not distinguish between Einstein's
special relativity and \dSR. Namely, all experiments that prove
Einstein's special relativity cannot exclude the \dSR.}

The interval and thus light-cone can be well defined as the
inverse Wick rotation counterparts of (\ref{AB}) and (\ref{ABL}).

\subsection{Two kinds of simultaneity and closed \dS-cosmos }

Different from Einstein's special relativity, there are two kinds
of simultaneity, and there is a relation between them reflecting
the cosmological significance of \dS\, \SR.

The first is of the Beltrami-time $t=x^0/c$ for the experiments of
inertial-observers with the principle of relativity. {Similar to
Einstein's \SR, one can define that two events $A$ and $B$ are
simultaneous if and only if the Beltrami-time coordinate $x^0$ for
the two events
are same, %i.e.
\be\nno %
a^0:=x^0(A) =x^0(B)=:b^0. %
\ee
It is with respect to this simultaneity that free particles move
along straight lines with uniform velocities. This simultaneity
defines a 3+1 decomposition of the Beltrami metric in one patch
\cite{BdS}.}

The second simultaneity is for the proper time $\tau$ of clocks at
rest in Beltrami coordinates {with the spatial Beltrami coordinates
$x^i={const}$. Namely, all events are simultaneous if and only if
they correspond the same $\tau$.}

The two time-scales are related by
\begin{eqnarray}\label{ptime}
\tau=R \sinh^{-1} (R^{-1}\sigma^{-1/2}(x)ct).
\end{eqnarray}
If $\tau$ is chosen as cosmic-time by comoving observers with
cosmological principle, the Beltrami-metric becomes its
Robertson-Walker counterpart \cite{BdS}
\be\label{dsRW}
ds^2=d\tau^2-dl^2=d\tau^2-\cosh^2( R^{-1}\tau) dl_{0}^2,%
\ee
with $dl_{0}^2$ is a 3-dimensional
Beltrami-metric on an $S^3$ with radius $R$. It is an `empty'
accelerated expanding cosmological model with a slightly closed
cosmos of order $R$.

{Since the Beltrami coordinates are essential for the principle of
relativity, the physical simultaneity corresponding to comoving
observations should be related to the Beltrami coordinate systems in
a most natural and simplest way. The family of `static' observers
with spatial coordinates $x^i={consts.}$ in the Beltrami coordinates
should be regarded as the comoving ones with respect to the
corresponding proper-time simultaneity. In fact, the relevant kind
of comoving coordinates of (\ref{dsRW}) are indeed most natural and
simplest in comparison with other kinds of comoving coordinates
having  flat or open 3-dimensional cosmos, respectively.}

If we take $R^2=3\Lambda^{-1}$, our universe is then asymptotic to
the closed \dS-cosmos (\ref{dsRW}). This is a prediction different
from standard cosmological models in \GR, in which there is an input
parameter $k$ to characterize whether the universe is closed or not.
{This prediction does not conflict with the results of WMAP and SDSS
\cite{WMAP}\footnote{The three year observations of WMAP (Table 12
of the last reference in \cite{WMAP}) show that the central values
of $\Omega_k$ systematically lean to negative and the error bar for
1$\sigma$ is entirely in the region $\Omega_k<0$ for the data set
WMAP+SDSS LRG.} and can be further checked}.

It is important that the \dS-group as a maximum symmetry ensures
that in  \dS-space there are the principle of relativity, the
cosmological principle of \dS-invariance and their relation as well.
In \dS-space there should be a type of {\it inertial-comoving
observers} having a kind of two-time-scale clocks, measuring  the
Beltrami-time and the cosmic-time. This reflects that there is an
important linking the principle of relativity with the cosmological
principle of \dS-invariance. Thus, what should be done for those
inertial-comoving observers is merely to switch their timers from
cosmic-time back to Beltrami-time according to their relation and
vice versa. Namely, once the observers would carry on the
experiments in their laboratories, they should take their timers
switching on Beltrami-time scale and off the cosmic-time scale so as
to act as inertial observers and all observations are of inertia,
while when they would take cosmic-observations on the distant stars
and the cosmic effects other than the cosmological constant as test
objects they may just switch off the Beltrami-time scale and on the
cosmic-time scale again, then they should act as a kind of comoving
observers as they hope.

Similar issues hold for the \AdS-space, too.

\subsection{On thermodynamical properties}

In ordinary approach to \dS-space in \GR, there should be the
Hawking-temperature and entropy at the horizon \cite{GH}. This
leads to one of the \dS-puzzles: Why does \dS-space of constant
curvature look like a black hole and what is the statistical
origin of the entropy?

In the viewpoint of \dS\, special relativity, however, there is a
different explanation \cite{T}. From Eq. (\ref{ptime}), it is easy
to see that for the imaginary Beltrami-time and the proper-time,
there is no periodicity for the former since both Beltrami-time axis
and its imaginary counterpart are straight-lines without coordinate
acceleration, but there is such a period in the imaginary
proper-time that is inversely proportional to the
Hawking-temperature ${c\hbar/(2\pi R k_B)}$ at the horizon. If the
temperature Green's function can still be applied here, this should
indicate that for the horizon in Beltrami coordinates it is at
zero-temperature and that there is no need to introduce entropy. In
addition, {in \dS\, special relativity, the simultaneity in an
inertial frame is defined by its Beltrami-time. A `test' mass moving
along any world line with the Beltrami coordinates $x^i=const$ is of
inertia and has vanishing {\it coordinate} velocity. Its 4-proper
acceleration, 3-(coordinate) acceleration, 4-(coordinate)
acceleration, and the relative (coordinate) acceleration between the
two nearby observers are all zero. Thus, there is no force needed
for an inertial observer to hold a `test' mass in place when it
tends to the event horizon. Namely, there is no `surface gravity' on
the event horizon in \dS-spacetime in the viewpoint of \dS\, special
relativity \cite{T}.

On the other hand, it can be shown that the non-vanishing surface
gravity on \dS-horizon given in \GR\ is actually a kind of {\it
inertial force}, which leads to the departure from an inertial
motion. For example, the observer near the \dS-horizon static with
respect to the Killing time is of non-inertia.} Therefore, although
there are Hawking-temperature and entropy $S=\pi R^2{c^3k_B/G\hbar}$
at the horizon in other coordinate-systems such as the static and
the Robertson-Walker-like ones (\ref{dsRW}), they are not caused by
gravity but by non-inertial motions. This is also in analogy with
the relation between Einstein's special relativity in \Mink-space
and the horizon in Rindler-coordinates. The temperature at the
Rindler-horizon is caused by non-inertial motion rather than gravity
\cite{T}.

%%%%%%%%%%%%%%%%%%%%%%%%%%%%%%%%%%%%%%%%%%%%%%%%
%
%              QST model
%
%%%%%%%%%%%%%%%%%%%%%%%%%%%%%%%%%%%%%%%%%%%%%%%%

\section{Snyder's Quantized Space-time and  DSR}

Snyder considered the homogenous quadratic form %
\be\label{snyder}%
-\eta^2=\eta^2_0-\eta^2_1-\eta^2_2-\eta^2_3-\eta^2_4
:=\eta^{AB}\eta_A^{}\eta_B^{} <0,
\ee%
partially inspired by Pauli. It is a model via homogeneous
(projective) coordinates of a 4-d {momentum} space of constant
curvature, a \dS-space of momenta. Snyder's model of (\ref{snyder})
may be reviewed as a \dS-hyperboloid (\ref{qiu2}) in 5-d space of
momenta, the inverse Wick rotation of (\ref{4s}), if we identify
$\eta_A$ with $(\al/l) \xi_A$ with a common factor $ (\al/l) \neq 0$
where $\al$ is near or equal to the Planck momentum, proportional to
the inverse of the parameter $a$ in Snyder's model:
\be\label{snydera}%
H_\al&:&\eta^{AB}\eta_{A}\eta_{B}=-\alpha^2,\\
&&ds_p^2=\eta^{A B} d\eta_A d\eta_B, \label{dsp} \\
\partial_P H_\al&:&\eta^{AB}\eta_{A}\eta_{B}=0.
\ee%
It should be pointed out that we may consider a massive `particle'
with a 5-momentum {$P_A=m_l{c}\eta_{AB}\frac{d\xi^B}{ds}$} moving
on the \dS-hyperboloid embedded in a 5-dimensional \Mink-space
\HlsM\, (\ref{qiu2}). Then, Snyder's condition (\ref{snyder}) or
(\ref{snydera}) of momenta indicate that it {is} satisfied by the
particle's 5-momentum $P_A$ if the particle could be a `tachyon'
with $m_l^2<0$:
\be\label{tchn}%
\eta^{AB}P_AP_B=m_l^2 c^2<0.
\ee%
In fact, this is a general issue not only for Snyder's model, but
also for other DSR models, which may be transformed from Snyder's
one (see, e.g. \cite{DSRdS}).

Further, a Beltrami model of \dS-space of momenta may also be set
up on a space of momenta:
\be\label{BdSp}%
p_0:=\alpha\frac{\eta_0}{\eta_4}=\alpha\frac{\xi^0}{\xi^4},
\quad p_i
:=\alpha\frac{\eta_i}{\eta_4}=\alpha\frac{\xi^i}{\xi^4}.
\ee%

Quantum mechanically according to Snyder, in this `momentum picture'
the operators for the time-coordinate $\hat t$ and the
space-coordinates $\hat x^i$ should be given by
\be\label{xt}%
\hat x^i&:=&i (\frac{\partial}{\partial p^i}+\al^{-2}p^i
p^\mu\frac{\partial}{\partial p^\mu}),\\\nno%
\hat x^0&:=&i (\frac{\partial}{\partial p^0}-\al^{-2}p^0
p^\mu\frac{\partial}{\partial p^\mu}), \quad\hat x^0=c\hat t,
\quad p^\mu=\eta^{\mu\nu}p_\nu.
\ee%
They form an $SO(1,4)$ algebra together with the `boost' $\hat
M^i=\hat x^i p^0+\hat x^0p^i$ and `3-angular
momentum' $\hat L^i= \frac 1 2 \eps ^{i}_{~jk}(\hat x^j p^k-\hat x^k
p^j)$
\be\label{so14m}%
&&[\hat x^i, \hat x^j]={i\al^{-2}}\epsilon^{ij}_{~~k}\hat
L^k,~
[\hat x^0, \hat x^i ]={i\al^{-2}}\hat M_i, \\\nno%
&&[\hat L^i, \hat L^j]=\epsilon^{ij}_{~~k}\hat L^k, \qquad [\hat
M^i, \hat M^j]=\epsilon^{ij}_{~~k}\hat M^k,~~etc.
\ee%

It should be noted that since $p^\mu$ as inhomogeneous (projective)
coordinates or Beltrami coordinates $q^\mu$ in (\ref{q}) after an
inverse Wick rotation, one coordinate patch is not enough to cover
the \dS-space of momenta in Snyder's model\omits{ and DSR models}
\cite{BdS}. Since the 4-d projective space $RP^4$ is not orientable,
in order to preserve the orientation the antipodal identification
should not be taken. In Snyder's model, the operators $\hat x^\mu$
are just four generators (\ref{generator}) of the \dS-algebra and
$\hat L^i, \hat M^i$ are just the rest six generators $\hat
L_{\mu\nu}$ in (\ref{Lorentz}) of Lorentz algebra $so(1,3)$.
Actually, the algebra (\ref{so14m}) is the same as (\ref{so14}) in
the space of momenta.

Similar to Snyder's model, a quantized space-time in \AdS-space of
momenta can also be constructed.

As mentioned in Introduction, most DSR models with
$\kappa$-Poincar\'e algebra can be realized geometrically by means
of other kinds of coordinate systems on \dS/\AdS-space of momenta
\cite{DSRdS} other than the inhomogeneous projective coordinates
used by Snyder's model in \dS-space of momenta. Thus, the relation
between these Snyder-like models and DSR (with $\kappa$-Poincar\'e
algebra) may be {given based on the coordinate transformations} on
4-d \dS/\AdS-space of momenta.

It is important that in these models after an inverse Wick
rotation the inverses of {ratios} in (\ref{um}) become `group
velocity' components of some `wave-packets':
\be\label{gv}%
\frac{\partial E}{\partial p^i}=consts. \quad E=p^0 c.
\ee%
Thus, there is a kind of uniform motions with component `group
velocity' or a {\it law of inertia-like} hidden in these models. In
particular, when the correspondences of $\beta^i$ in (\ref{sl})
vanish, the `group velocity' of a `wave-packet' coincides with its
`phase velocity'. This is similar to the case for a light pulse
propagating in vacuum \Mink-spacetime.

On the other hand, the treatment parallel to the one on the
\dS-space in general relativity will lead to the conclusions that
there are `temperature' $\tilde{T}_p$ and `entropy' $\tilde{S}_p$
at the horizon in \dS-space of momenta. It is unclear what sense
$\tilde T_p$ makes in DSR models other than Snyder's.  However,
the treatment parallel to the one in the viewpoint of \dS\,
special relativity gives rise to the zero-`temperature' $\tilde
T_p$ without `entropy'.

%%%%%%%%%%%%%%%%%%%%%%%%%%%%%%%%%%%%%%%%%%%%%%%%%%
%
%              Duality
%
%%%%%%%%%%%%%%%%%%%%%%%%%%%%%%%%%%%%%%%%%%%%%%%%%%

\section{Duality between Planck Scale and Cosmological Constant }

\subsection{A one-to-one correspondence between Snyder's model and  \dS\ special relativity}

It is important to see that there is  an interesting and important
dual one-to-one correspondence in \BdS\ between Snyder's model and
\dS\, special relativity as is shown in Table 1.

\bc
\begin{tabular}{lcl}
Table 1.  Dual correspondence of&&\hskip -1mm
Snyder's %\\
model and \dS\, special relativity \\
\hline
{\quad Snyder's model} && {~ \dS\, special relativity}\\
\hline
momentum `picture' && coordinate `picture'\\
\BdS-space of momenta && \BdS-spacetime\\
radius $\al\sim $ Planck scale && radius $R \sim$ cosmic  radius \\
constant `group velocity'&& constant 3-velocity\\
quantized space-time && `quantized' momenta \\
$\hat x^i, \hat t$&&$\hat p_i, \hat E$\\
$ \tilde{T}_p=0$ {\rm without} $\tilde{S}_p$ && $T=0$ {\rm
without}
$S$\\
\hline
\end{tabular}
\ec

The dual one-to-one correspondence should not be considered to
happen accidentally. Since the Planck length and the cosmological
constant provide a UV and an IR scale, respectively, this
correspondence is a kind of the UV-IR connection and should reflect
some dual relation between the physics at these two scales.

\subsection{A minimum uncertainty-like argument and the dual correspondence }

In fact, there is a minimum uncertainty-like argument that may
indicate why there should be a one-to-one correspondence between
Snyder's model and \dS\ special relativity.

Quantum mechanically, the coordinates and momenta cannot be
determined exactly at the same time if there is an uncertainty
principle in the embedded space
\be \label{upr}
\Dl \xi^I \Dl \eta_I^{} \geq \frac \hbar 2, %
 \ee
where $I=1, \cdots, 4$ and the sum over $I$ is not taken. It
should be mentioned that here we simply employ the same notation
of some observable for the expectation value of its operator over
wave function in quantum mechanics.

Limited on the hyperboloid in embedded space, i. e. $\Dl \xi^I \leq
R$, and suppose that the momentum $\eta_I^{}$ conjugate to $\xi^I$
also takes values on a hyperboloid. Then, $\Dl \eta_I^{}\leq \alpha$
and the minimum uncertainty relation implies $R \alpha \sim \hbar$.
{Note that in this subsection $R$ and $\al$ are free radius
parameters for \dS-spacetime and \dS-space of momenta,
respectively.} When the size of hyperboloid in the space of
coordinates is Planck length, namely,
\be
\eta_{AB}\xi^A \xi^B = - \ell_P^2 = -G\hbar c^{-3},
\ee
the hyperboloid in the space of momenta then has Planck scale,
\be \eta^{AB}\eta_A \eta_B =  - E_P^2/c^2 =  -\hbar
c^3/G <0, \ee
which is equivalent to the Snyder's relation (\ref{snyder}).  On the contrary, when the scale of hyperboloid in the space of momenta is %
\be
\eta^{AB}\eta_A \eta_B =- \frac  {\La\hbar^2} 3,
\ee
then we have relation (\ref{qiu2}). Therefore, the minimum
uncertainty-like argument indicates a kind of the UV-IR connection
and the one-to-one correspondence listed in the Table 1 should
reflect some duality between the physics at these two scales.

This argument may further be supported by a `tachyon' dynamics as
follows.

Suppose there is a free particle with mass $m_R$ moving along a
time-like curve $C(\varsigma)$ on \dS-hyperboloid \HrsM\, with an action%
\be\label{act}%
 S={\frac 1 c}\int_{C(\varsigma)} d{\varsigma} L=m_R{c}\int_{C(\varsigma)} d\varsigma \sqrt{\eta_{AB} \dot{\xi}^A\dot{\xi}^B},
 \qquad \dot{\xi}^A=\frac{d \xi^A}{d\varsigma}, %
\ee%
where $\varsigma$, dimension of length, is an affine parameter
for the curve $C(\varsigma)$ and it may be taken as $\varsigma=s$,
and $L$ the Lagrangian.

A (vertical) variation and the variational principle lead to the
Euler-Lagrangian equation equivalent to (\ref{angular5a}) so long as
the affine parameter $\varsigma$ being taken as $s$. And  horizontal
variations as Lie derivatives with respect to those Killing vectors
show that the 5-d angular momentum is Noether's charges. We may also
introduce an action with a Lagrangian multiplier for the embedding
condition (\ref{qiu2}) and the results are the same.

In order to find its phase space, we may take a Legendre
transformation to get a Hamiltonian and suppose the basic Poisson
brackets as usual. Alternatively, we may also take a `vertical'
differential $d_v$ of the Lagrangian and from $d_v^2L=0$, we may
further read off the symplectic form and canonical variables (see,
e.g., \cite{gwjmp}). Here $d_v$ is nilpotent and anti-commutative
with $d_\varsigma:=d$ above or it
 may also be regarded as a nilpotent `external' variation.
 Thus, we have
\be%\begin{align}
d_v {\it L}={\cal E}+\frac{d}{d\varsigma}\pmb{\theta},
\ee%
where
\be%
{\cal E} = -\eta_{AB}{ m_R^{} c}(\ddot{\xi}^A-R^{-2} \xi^A)d_v \xi^B %
\ee is called Euler-Lagrange 1-form \cite{gwjmp},
\be%
 \pmb{\theta} = m_R^{}{ c} \eta_{AB} \dot{\xi}^A d_v \xi^B
\ee
the symplectic 1-form, {and $\zeta$} the parameter in the action
(\ref{act}). Then, from the nilpotency of $d_v$, it follows a
necessary and sufficient condition for the preserving of a symplectic form%
\be 0=d_v^2{\it L}=\frac{d}{d\varsigma}\pmb{\omega}+d_v{\cal E},%
\ee%
where the symplectic structure may be given in the canonical form as
\be\label{syms}%
\pmb{\omega}=d_v \pmb{\theta}=d_v P_B \wedge d_v \xi^B, \quad
P_B:=m_R^{}{ c} \eta_{AB}^{}\frac{d{\xi}^A}{ds}, ~~(\varsigma=s).
\ee%
Here $P_B$ are canonical momentum for the canonical formalism. This
symplectic structure (\ref{syms}) shows that $(\xi^A, P_B)$ should
be a set of canonical variables on the phase space with basic
Poisson brackets given by the symplectic structure:
\be\label{psn}%
 \{\xi^A, P_B\}=\delta^A_B.%
\ee%
Then, the conserved quantities $L^{AB}$ for the particle form an
$\mathfrak{so}(1,4)$ algebra in Poisson bracket:
\be \{L^{AB},L^{CD}\} &=&\eta^{AC}
L^{BD}-\eta^{AD} L^{BC}%\nno \\
+\eta^{BD}
L^{AC}- \eta^{BC} L^{AD}. \label{eq:na} %
\ee%
It should be noted that the canonical momenta $P_B$ for the particle
automatically satisfy (\ref{tchn}), if we would identify $P_A$ with
$\eta_A^{}$. This should imply that the particle be a `tachyon' on
\HrsM.

Quantum mechanically, the canonical variables $(\xi^A, P_B)$ with
their Poisson brackets (\ref{psn}) become
\be\label{qbrct}%
\hat P_B:=-i{ \hbar } \frac{\partial}{\partial \xi^B}, \qquad
[\xi^A, \hat P_B]=-i {\hbar}\delta^A_B,
\ee%
in a coordinate picture. And the conserved quantities $L^{AB}$
become the operators ${\hat{\cal L}}_{AB}$ of (\ref{Generator}),
which lead to a \dS-algebra in Lie bracket. We may also write them
in a momentum picture.

Further, on a `simultaneous' 3-hypersurface $\xi^0=const.$, we may
have an uncertainty relation like (\ref{upr})
for such a `tachyon' quantum mechanically. %
Thus, such a `tachyon' dynamics may support why there is a
one-to-one dual correspondence of Snyder's model and \dS\, \SR.

We may also study such a `tachyon' dynamics directly in a
\BdS-model. This should lead to the operators  $\hat t$ and $\hat
x^i$ in (\ref{xt}) in a momentum picture given by Snyder in his
model.

\subsection{The duality between Planck scale and cosmological constant}

%%%%%%%%%%%%%%%%%%%%%%%%%%%%%%%%%%%%%%%%%%%%%%

It is clear that both Snyder's model and \dS\, special relativity
may deal with the motion of relativistic particles. In \dS\, special
relativity, the momenta of a particle are quantized and
noncommutative, while in Snyder's model, the coordinates of a
particle are quantized and noncommutative. These are listed in Table
1 for the one-to-one correspondence between them. As was mentioned
at beginning, however, the dimensionless constant $g {\sim}
\ell_P/R$ in (\ref{g}) contains Newton's gravitational constant and
thus should describe some gravity. Therefore, we may make such a
conjecture: {\it the physics at such two scales should be dual to
each other in some `phase' space and there is in-between gravity of
local \dS-invariance characterized by $g$.}

It is interesting to notice that $ g^2$ is in the same order of
difference between $\Lambda$ and the theoretical quantum `vacuum
energy', there is no longer the puzzle in the viewpoint of the \dS\,
special relativity and gravity with local \dS-invariance.

However, since $\Lambda$ is a fundamental constant as  $c, G$ and
$\hbar$, a further question should be: What is the origin of the
dimensionless constant $g$? Is it calculable?

It is just the first question of the `top ten' {\cite{top10}}: `Are
all the (measurable) dimensionless parameters that characterize the
physical universe calculable in principle or are some merely
determined by historical or quantum mechanical accident and
incalculable?'

It is important to note that there are some hints on the answer for
this dimensionless constant $g$. First, since among 4-d Euclid,
Riemann and Lobachevski spaces there is only the Riemann-sphere with
non-vanishing 4-d topological number. Thus, if there is a quantum
tunneling scenario for the Riemann-sphere ${ S}^4$ as an instanton
of gravity to the \dS-space, this can explain why the cosmological
constant should be positive, i.e. $\Lambda>0$. We should show in the
next section that in a simple \dS-Lorentz model of \dS-gravity this
is just the case.

Further, if the action of the \dS-gravity is of the Yang-Mills
type, then its Euclidean version is of a non-Abelian type with
local $SO(5)$ symmetry. Thus, due to the asymptotic freedom
mechanism, the dimensionless coupling constant, say $g$, should be
running and approaching to zero as the momentum tends to
infinity. However, for the case of gravity, the momentum could not
{tend} to infinity but the Planck scale so that the Euclidean
counterpart of the dimensionless coupling constant should be very
tiny.

Needless to say, to completely explain this problem is still a
long way to go.

\section{How to Describe Gravity Characterized by $g$?}

\subsection{Gravity as the localization of relativity
with a gauge-like dynamics}

In order to explain how to describe gravity, we should first recall
Einstein's equivalence principle, which says roughly that the laws
of physics in a freely falling (nonrotating) lift are the same as
those in inertial frames in a flat spacetime.  Some scholars even
say that the spacetime for a freely falling observer will be that of
Einstein's special relativity (known as \Mink-spacetime)
\cite{Peacock}. It is well known that in a \Mink-spacetime the full
Poincar\'e symmetry should hold. However, in Einstein's general
relativity only the local Lorentz symmetry is preserved in a local
frames \cite{Cartan,EC,held}. One may establish the gauge theory of
local Poincar\'e symmetry for gravity \cite{EC,held,An}, which may
be regarded as the localization of Einstein's special relativity
with full Poincar\'e symmetry. Similarly, one may established the
gauge theory of local \dS/\AdS-symmetry based on the localization of
\dS/\AdS\, special relativity, respectively. In this sense, the
gauge theory of gravity should be based on the localization of
corresponding \SR\ with full symmetry.

As for the dynamics of gravity, we may expect that the gravity is
governed by a gauge-like dynamics with the dimensionless coupling
constant $g$. This is also in consistent with the localization of
special relativity. Of course, correct equations should pass
observation tests for \GR\ at least.

How to properly describe gravity based on the localization of
\dS\, special relativity and governed by a gauge-like dynamics is
still under investigation. In the following we shall show that
these points have been indicated by a simple model of \dS-gravity
\cite{dSG, uml, T77,QG} on a kind of umbilical manifolds of
local \dS-invariance in a special gauge \cite{uml}.

\subsection{Localization of \dS-hyperboloid on umbilical manifolds}

Simply speaking, the spacetimes with gravity of local
\dS-invariance may be described as a kind of $(3+1)$-dimensional
umbilical manifolds ${\cal M}^{1,3}:={\cal H}^{1,3}$ as
sub-manifolds of $(4+1)$-dimensional Riemann-Cartan manifolds
${\cal M}^{1,4}$. In surface theory \cite{GW}, a surface is
umbilical if the normal curvatures at its each point are a
constant. A sphere of radius $R$, $S^2 \subset E^3$, is such an
umbilical surface, on
which %
\be\label{umbilic}%
g_{\mu\nu}=Rb_{\mu\nu},%
\ee%.
holds at each point, where $g_{\mu\nu}, b_{\mu\nu}$ are the first
and second fundamental form of the manifold, respectively.  The
localization or the fibration of $S^2\subset E^3$ may lead to a
2-dimensional umbilical manifold ${\cal S}^2$ with all points
being umbilical as a submanifold of a 3-dimensional manifold
${\cal E}^3$. It is in this way the localization of a
\dS-hyperboloid \HrsM\, could be realized \cite{uml}.

Let us now illustrate how to construct such an ${\cal
M}^{1,3}:={\cal H}^{1,3} \subset {\cal M}^{1,4}$.

Suppose ${\cal H}^{1,3}$ is also of Riemann-Cartan with signature
$-2$. Thus, for any given  point $\forall p \in {\cal H}^{1,3}$,
there is a local \Mink-space as the tangent space at the point,
$T_p({\cal H}^{1,3})$, and given a vector {$N_p=Rn_p$} of norm $R$
at the point with an $n_{p}$ as the unit base of space $N^1_p$
normal to $T_p({\cal H}^{1,3})$ with a metric of \dS-signature in
${\cal M}^{1,4}$. Then the space $T_p \times {N }^1_p \cong
M^{1,4}_p$ is tangent to ${\cal M}^{1,4}$ at the point. Thus, under
local \dS-transformations on $T_p \times {N }^1_p \cong M^{1,4}_p$
there is a local hyperboloid structure $H_R \subset M^{1,4}_p$
isomorphic to the \dS-hyperboloid \HrsM\, in (\ref{qiu2}) at the
point $p$ as long as $R{n}_p=-{\bf r}_p$ is taken, where ${\bf r}_p$
is the radius vector. In fact, all these points consist of the
umbilical manifold ${\cal M}^{1,3}:={\cal H}^{1,3} \subset {\cal
M}^{1,4}$.

This construction can also be given in an opposite manner. Suppose
there is a local \HrsM\, anywhere and anytime tangent to the ${\cal
M}^{1,4}$ such that at a point $p \in {\cal H}^{1,3}$, the radius
vector ${\bf r}_p$ with norm $R$ of the \HrsM\, is oppositely normal
to the tangent \Mink-space of ${\cal H}^{1,3}$, i.e. ${\bf r}_p=-{
N}_p$, at the point. {Since this local \Mink-space is also tangent
to the \HrsM\, at the point so that the point is an umbilical point
in ${\cal M}^{1,4}$. Thus, ${\cal H}^{1,3}$ consists of all these
umbilical points and is a sub-manifold of the ${\cal M}^{1,4}$,
i.e., ${\cal H}^{1,3} \subset {\cal M}^{1,4}$. Such a kind of
Riemann-Cartan manifolds ${\cal H}^{1,3}$ are called umbilical
manifolds with an umbilical structure of \HrsM\, anywhere and
anytime.}

Therefore, on the co-tangent space $T_p^*$ at the point $p \in
{\cal
H}^{1,3}$ there is a Lorentz frame 1-form:%
\be\label{Lframe}%
\theta^b=e^b_\mu dx^\mu, ~~\theta^b(\partial_\mu)=e^b_\mu; \quad
e^a_\mu e^\mu_b=\delta^a_b, ~~~e^a_\mu e_a^\nu=\delta^\nu_\mu; %
\ee%
with respect to a Lorentz inner product:%
\be\label{Lpro}%
<\partial_\mu,\partial_\nu>=g_{\mu\nu}, ~~<e_a, e_b>=\eta_{ab},
~~\eta_{ab}={ {\rm diag}}(1,-1,-1,-1). %
\ee%
Here, $\partial_\mu$ {is} the {coordinate} base of the tangent space $T_p$.
The line-element on ${\cal H}^{1,3}$
can be expressed as%
\be%
ds^2=g_{\mu\nu}dx^\mu dx^\nu=\eta_{ab}\theta^a\theta^b,
\quad~g_{\mu\nu}=\eta_{ab}e^a_\mu e^b_\nu.%
\ee%

There is a Lorentz covariant derivative a la Cartan:
\be%
 \nabla_{e_a}e_b=\theta^c_{~b}(e_a)e_c; \quad %
 \theta^a_{~b}=B^a_{~b \mu}dx^\mu,\quad~\theta^a_{~b}(\partial_\mu)=B^a_{~b\mu}. %
\ee%
$B^{ab}_{~\mu}=\eta^{bc}B^a_{~c\mu}\in \mathfrak{so}(1,3)$ are
connection coefficients of the Lorentz connection 1-form
$\theta^{ab}=\eta^{bc}\theta^a_{~c}$.
The torsion and curvature can be defined as%
\be\nno%
\Omega^a&=&d\theta^a+\theta^a_{~b}
\wedge\theta^b=\frac{1}{2}T^a_{~\mu\nu}dx^\mu\wedge
dx^\nu,\\
\label{T2form}&&T^a_{~\mu\nu}=\partial_\mu e^a_\nu-\partial_\nu e^a_
\mu+B^a_{~c \mu}e^c_\nu-B^a_{~c
\nu}e^c_\mu;\\ \nno%
\Omega^a_{~b}&=&d\theta^a_{~b}+\theta^a_{~c}\wedge\theta^c_{~b}=\frac{1}{2}F^a_{~b
\mu\nu}dx^\mu\wedge dx^\nu,\\
\label{F2form}&&F^a_{~b \mu\nu}=\partial_\mu B^a_{~b\nu}
-\partial_\nu B^a_{~b\mu}+B^a_{~c\mu}B^c_{~b
\nu}-B^a_{~c\nu}B^c_{~b\mu}.%
\ee%
They satisfy corresponding Bianchi identities.

It is easy to get a metric compatible affine connection
$\Gamma^\la_{~\mu\nu}$ from the requirement
\be \label{De=0}%
\omits{g_{\mu\nu {;}\la}=0, ~\Leftrightarrow~}e^a_{~\mu {;}
\nu}=\partial_\nu
e^a_{~\mu}-\Gamma^\la_{~\mu\nu}e^a_{~\la}+B^a_{~c\nu}e^c_{~\mu}=0,
\ee%
where $;$ denotes the covariant derivative with respect to the
affine connection $\Gamma^\la_{~\mu\nu}$ for spacetime indexes and
the spin connection $B^a_{~c\nu}$ for Lorentz-frame indexes.

As was just mentioned, at the point $p \in {\cal H}^{1,3}$, there
are a space $N^1_p$ and its dual ${N^1_p}^*$ normal to ${\cal
H}^{1,3}$ with a normal vector $n$ and its dual $ \nu$ in
$T_p({\cal M}^{1,4})$ and $T^*_p({\cal M}^{1,4})$, respectively.
Namely, {both} $\{\partial_\mu, n; dx^\la,\nu\}$ and $\{e_a,n;
\theta^b, \nu\}$ span $M_p^{1,4}=T_p^{1,3}\times N_p^1$ and
${M_p^{1,4}}^*={T_p^{1,3}}^*\times {N_p^1}^*$. Let these
bases satisfy the following conditions in addition to (\ref{Lpro})%
\be\label{dSpro}%
dx^\la(n)=\theta^b(n)=0,&&\nu(\partial_\mu)=\nu(e_a)=0,~~~n(\nu)=1;\\%\nno
<e_a,n>{ =<\r_\mu,n>}=0,&&<n,n>=-1.%
\ee %
Then, the \dS-Lorentz base $\{\hat{E}_A\}$ and their dual
$\{\hat\Theta^B\}$ {($A, B = 0,\cdots, 4$)} can be defined as: %
\be\label{dSL}%
\{\hat{E}_A\}=\{e_a, n\},~~\{\hat{\Theta}^B\}=\{\theta^b, \nu\}.%
\ee%
And (\ref{Lpro}) and (\ref{dSpro}) can be expressed as%
\be\label{dSLpro}%
\hat\Theta^B(\hat E_A)=\delta^B_A, ~~<\hat E_A,
\hat E_B>=\eta_{A B}^{}={\rm diag}(1,-1,-1,-1,-1).%
\ee%
Introduce a normal vector $N=Rn$ with norm $R$:%
\be\label{N}%
N=Rn=\hat \xi^A \hat E_A, ~~(\hat \xi^A)=(0,0,0,0,R), ~~<N,N>=-R^2.%
\ee%
For  the \dS-Lorentz base, there are %
\be\label{LocalS}%
g_{\mu\nu}=\eta_{AB}\hat E^A_\mu \hat
E^B_\nu,~~\eta_{AB}\hat\xi^A\hat
E^b_\mu=0,~~\eta_{AB}\hat \xi^A\hat\xi^B=-R^2,%
 \ee%
where %
\be%
\hat E^A_\mu=\hat \Theta^A(\partial_\mu), ~~\{\hat
E^A_\mu\}=\{e^a_\mu, 0\}. %
\ee%

The transformations, which maps $M^{1,4}_p$ to itself and
preserves the inner product, are
\be%
\hat E_A\rightarrow E_A={(S^t)}^{\ B}_A \hat E_B, ~~\hat
\Theta^A\rightarrow
\Theta={(({ S^t})^{-1})}^{\ A}_B \hat \Theta^B, ~~S^t{J}S={J},%
\ee%
where ${J}=(\eta_{AB}^{}),~S = (S^A_{\ B})\in SO(1,4)$, the
superscript $t$  denotes the transpose as in (\ref{dst}). The
transformed base is defined as the \dS-base and its dual
$E_A, \Theta^B$, respectively: %
\be\label{dSB}%
\Theta^A(E_B)=\delta^A_B,&&\Theta^A(\partial_\mu)=E^A_\mu,\quad~
<E_A,E_B>=\eta_{AB}.\\\label{LocalS1} g_{\mu\nu}=\eta_{AB}E^A_\mu
E^B_\nu,&&\eta_{AB}\xi^A
E^B_
\mu=0,\quad~\eta_{AB}\xi^A\xi^B=-R^2,%
\ee%
where $E^B_\mu$ are the \dS-frame coefficients. Obviously, these
formulas reflect the local \dS-invariance on ${\cal H}^{1,3}$ and
Eqs. in (\ref{LocalS1}) show that there is a local 4-dimensional
hyperboloid $H^{1,3}_p\subset M^{1,4}_p$ tangent to ${\cal
H}^{1,3}$ at the point $p$. Thus, (\ref{LocalS1}) may be called
{\it the local \dS-hyperboloid condition}.

Now  the \dS-covariant derivative a la Cartan can be introduced%
\be\label{dSCD}%
\hat \nabla_{E_A}E_B=\Theta^C_{~B}(E_A)E_C.%
\ee%
$\Theta^{AB}=\eta^{BC}\Theta^A_{~C}\in \mathfrak{so}(1,4)$ is the
\dS-connection 1-form. In the local
coordinate chart $\{x^\mu\}$ on ${\cal H}^{1,3}$,%
\be\label{dSconnection}%
\hat \nabla_{\partial_\mu}E_B=\Theta^C_{~B}(\partial_\mu)E_C={B}^C_{~B\mu}E_C,
\quad {\hat \nabla_{n}E_B=\Theta^C_{~B}(n)E_C = {B}^C_{~B4}E_C,} %
\ee %
${B}^{AB}_{~~\mu}=\eta^{BC}{B}^A_{~C\mu}$ {\it and ${B}^{AB}_{~~4}=\eta^{BC}{B}^A_{~C4}$} denote the
\dS-connection coefficients on ${\cal H}^{1,3}$. There are also
the \dS-torsion ${\bf \Omega}^A$, curvature 2-forms ${\bf
\Omega}^A_{~B}$ and  their Bianchi identities.

Importantly, in the light of  Gauss formula and Weingarten formula
in the theory of surfaces \cite{GW}, from the \dS-covariant
derivative of the \dS-Lorentz base (\ref{dSL}) with  properties of
$\theta^a, ~\theta^a_b$, it follows a
generalization of  Gauss formula and Weingarten formula%
\be\label{dSLGauge}%
\hat\nabla_{\partial_\mu}e_a=\theta^b_{~a}(\partial_\mu)e_b-b_{ab}\theta^b(\partial_\mu)
n,
~~\hat\nabla_{\partial_\mu}n=b^a_{~b}\theta^b(\partial_\mu)e_a.%
\ee%
Here, $b_{ab}$ denotes {the} second fundamental form of the
hypersurface.
Since ${\cal H}^{1,3}$ is supposed to be an umbilical
manifold, where every point satisfies the umbilical condition (\ref{umbilic}), %
these formulas read on ${\cal H}^{1,3}$%
\be\label{dSLGaugeS}%
\hat\nabla_{\partial_\mu}e_a=\theta^b_{~a}(\partial_\mu)e_b-R^{-1}\theta_a(\partial_\mu)n,
~~\hat\nabla_{\partial_\mu}n=R^{-1}\theta^a(\partial_\mu)e_a.%
\ee%

On the other hand, for the \dS-Lorentz base from
(\ref{dSCD}) there are%
\be\label{dSLCD}%
\hat\nabla_{\partial_\mu}e_a=\check
{\Theta}^b_{~a}(\partial_\mu)e_b+\check{\Theta}^4_{~a}(\partial_\mu)n,
~~\hat\nabla_{\partial_\mu}n=\check{\Theta}^a_{~4}(\partial_\mu)e_a,%
\ee%
where $\check{\Theta}$ denotes the \dS-connection ${\Theta}$ in
the \dS-Lorentz frame.

Comparing with (\ref{dSLGaugeS}), it follows%
\be\label{dSLconnect}%
\check{\Theta}^{ab}(\partial_\mu)={\theta}^{ab}(\partial_\mu)=B^{ab}_{~~{\bf \mu}},
&&~\check{\Theta}^{a{ 4}}(\partial_\mu)=R^{-1}{\theta}^{a{ 4}}(\partial_\mu)=R^{-1}e^a_\mu; \\\nno%
\check{ B}^{ab}_{~~{\mu}}=B^{ab}_{~~{ \mu}}, &&~\check{ B}^{a{4}}_{~~{\mu}}=R^{-1}e^a_\mu.%
\ee%
Namely, the \dS-connection in the \dS-Lorentz frame reads
\be\label{dSc}%
(\check {B}^{AB}_{~~{\mu}})=\left(
\begin{array}{cc}
B^{ab}_{~~{\mu}} & R^{-1} e^a_\mu\\
-R^{-1}e^b_\mu &0
\end{array}
\right ) \in \mathfrak{so}(1,4).
\ee%
This is just the connection introduced in
\cite{dSG,uml,T77,QG,Wise}. Here, it is recovered from the
umbilical manifolds with local \dS-invariance in the \dS-Lorentz
frame.

For the \dS-connection (\ref{dSc}), its curvature reads
\be\label{dSLF}%
{\check {\cal F}}_{\mu\nu}= ( \check{\cal F}^{AB}_{~~~\mu\nu})%
=\left(
\begin{array}{cc}
F^{ab}_{~~\mu\nu} + R^{-2}e^{ab}_{~~ \mu\nu} & R^{-1} T^a_{~\mu\nu}\\
-R^{-1}T^b_{~\mu\nu} &0
\end{array}
\right ) \in so(1,4),
\ee%
where $e^a_{~b\mu\nu}=e^a_\mu e_{b\nu}-e^a_\nu e_{b\mu},
e_{a\mu}=\eta_{ab}e^b_\mu$, $ F^{ab}_{~~ \mu\nu}$ and $
T^a_{~\mu\nu}$ are  curvature (\ref{F2form}) and torsion
(\ref{T2form}) of the Lorentz-connection.

\subsection{A simple model of \dS-gravity}
\label{sect:dS-gravity}

Now we consider the simple model of such \dS-gravitational fields
with a gauge-like action. The same \dS-connection with different
dynamics has also been explored in Ref. \cite{MM,SW,Wil,FS,Lec}.

The total action of the model with source may be taken as%
\be\label{S_t}%
S_{\rm T}=S_{\rm GYM}+S_{\mathrm M},%
\ee%
where $S_{\rm M}$ is the  action of the source with minimum
coupling, and $S_{\rm GYM}$ the gauge-like action in Lorentz gauge
of the model as follows \cite{dSG,T77,QG}:
\be\nno%
S_{\rm GYM}&=&\frac{{ \hbar}}{4g^2}\int_{\cal M}d^4 x e
{\bf Tr}_{dS}(\check{\cal F}_{\mu\nu}\check{\cal F}^{\mu\nu})\\
&=& -\int_{\cal M}d^4x e
\left[\frac{\hbar}{4g^2}F^{ab}_{~\mu\nu}F_{ab}^{~\mu\nu}-\chi(F-2\Lambda)
\right .\left. - \frac{\chi}{2}
T^a_{~\mu\nu}T_a^{~\mu\nu}\right].\label{GYM}
\ee%
Here, $e=\det(e^a_\mu)$, a dimensionless constant $g$ should be
introduced as usual in the gauge theory to describe the
self-interaction of the gauge field, $\chi$ a dimensional coupling
constant related to $g$ and $R$, and $F= \frac{1}{2}
F^{ab}_{~\mu\nu}e_{ab}^{~\mu\nu}$ the scalar curvature of the Cartan
connection, the same as the action in the Einstein-Cartan theory. In
order to make sense in comparing with the Einstein-Cartan theory, we
should take $R = (3/\Lambda)^{1/2}$, $\chi={ c^3}/({ 16}\pi G)$ and
$g^{-2} = 3\chi{ \hbar^{-1}}\Lambda^{-1}$.
$g^2$ defined here is the same order as the one
introduced in Eq.(\ref{g}) in the sense of the duality. This is why we have used the same symbol in
the different cases.

The field equations can be given via the variational principle with
respect to $e^a_{~\mu},B^{ab}_{~~\mu}$,\footnote{In what follows,
the unit of $c=1$ is used.}
\be\label{Geq2}%
T_{a~~||\nu}^{~\mu\nu } &-& F_{~a}^\mu+\frac{1}{2}F e_a^\mu -
\Lambda
e_a^\mu = 8\pi G( T_{{\rm M}a}^{~~\mu}+T_{{\rm G}a }^{~~\mu}), \\
\label{Geq2'}%
F_{ab~~||\nu}^{~~\mu\nu} &=& R^{-2}(16\pi G S^{\quad \mu}_{{\rm M}ab}+S^{\quad \mu}_{{\rm G}ab}).%
\ee
In Eqs.(\ref{Geq2}) and (\ref{Geq2'}), $||$ represents the covariant
derivative compatible with Christoffel symbol $\{^\mu_{\nu\ka}\}$
and spin connection $B^a_{\ b\mu}$,
$F_a^{~\mu}=-F_{ab}^{~~\mu\nu}e^b_\nu$, $F=F_a^{~\mu} e^a_\mu$,
\be
T_{{\rm M}a}^{~~\mu}&:=&-\d 1 e \d {\dl S_{\rm M}}{\dl e^a_\mu}, \\
\label{emG}%\nno
T_{{\rm G}a}^{~~\mu}&:=&g^{-2} T_{{\rm F}a}^{~~\mu}+2\chi T_{{\rm T}a}^{~~\mu},
\ee
are the tetrad form of the stress-energy tensor for matter and gravity, respectively, where
\be
\label{emF}%\nno
T_{{\rm F}a}^{~~\mu}&:=&-\frac{1}{4e} \frac{\delta} {\delta e^a_{\mu}}\int d^4x e {\rm Tr}(F_{\nu\ka}F^{\nu\ka}) \nno \\
&=& e_{a}^\ka {\rm Tr}(F^{\mu \la}F_{\ka \la})-\frac{1}{4}e_a^\mu
{\rm Tr}(F^{\la \si} F_{\la \si}) \ee
is the tetrad form of the stress-energy tensor for curvature and
\be\label{emT}%
T_{{\rm T}a}^{~~\mu}&:=&-\frac{1}{4e} \frac{\delta} {\delta e^a_{\mu}}\int d^4x e T^b_{\ \nu\ka}T_b^{\ \nu\ka}\nno \\
&=& e_a^\ka T_b^{~\mu\la}T^{b}_{~\ka\la}-\frac{1}{4}e_a^\mu
T_b^{~\la\si}T^b_{~\la\si}
\ee%
the tetrad form of the stress-energy tensor for torsion;
\be S_{{\rm M}ab}^{\quad \, \mu} =\d 1 {2\sqrt{-g}}\d {\dl S_{\rm
M} }{\dl B^{ab}_{\ \ \mu}} \ee and $S_{{\rm G}ab}^{\quad \,\mu }$
are spin currents for matter and gravity, respectively.
Especially, the spin current for gravity can be divided into two
parts,
\be\label{spG}%
S_{{\rm G}ab}^{\quad \, \mu}&=&S_{{\rm F}ab}^{\quad \, \mu}+2S_{{\rm T}ab}^{\quad \,\mu},
\ee
where
\be%
S_{{\rm F}ab}^{\quad \, \mu}&:=&\d 1 {2\sqrt{-g}}\d {\dl }{\dl
B^{ab}_{\ \
\mu}}\int d^4 x \sqrt{-g}F\nno \\
&=&{-}e^{~~\mu \nu}_{ab\ \ {||}\nu} = Y^\mu_{~\, \la\nu}
e_{ab}^{~~\la\nu}+Y ^\nu_{~\, \la\nu } e_{ab}^{~~\mu\la} \\
S_{{\rm T}ab}^{\quad \mu}&:=& \d 1 {2\sqrt{-g}}\d {\dl }{\dl B^{ab}_{\ \ \mu}}\d 1 4 \int d^4 x \sqrt{-g}T^c_{\ \nu\la}T_c^{\ \nu\la}\nno \\
&=&T_{[a}^{~\mu\la}e_{b]\la}^{}
\ee%
are the spin current for curvature $F$ and torsion $T$, respectively.

For the case of spinless matter and torsion-free gravity, the
curvature $F_{ab}^{~~\mu\nu}$ becomes the torsion-free curvature
${\cal R}_{ab}^{~~\mu\nu}$, the gravitational action
(\ref{GYM}) becomes%
\be%
S_{GYM}&=&\frac{1}{4g^2}\int_{\cal M}d^4x e
{\bf Tr}_{dS}({\cal R}_{\mu\nu}{\cal R}^{\mu\nu})\nno \\
&=& -\int_{\cal
M}d^4x
e \left[\frac{1}{4g^2}R^{ab}_{~\mu\nu}R_{ab}^{~\mu\nu}-\chi(R-2\Lambda)
\right],\qquad \label{rGYM}%
\ee%
and the field
equations (\ref{Geq2}) and (\ref{Geq2'})
become the Einstein-Yang equations \cite{gwz} with $\Lambda$-term%
\be%
&&{\cal R}_{~a}^\mu-\frac{1}{2}e_a^\mu {\cal R}+\Lambda
e_a^\mu=-8\pi G (T_{{\rm M}a}^{~~\mu}+ g^{-2} T_{{\rm R}a}^{~~\mu}), \label{Geq3}\\
&&{\cal R}_{ab~~||\nu}^{~~\mu\nu}=0.\label{Geq3'}
\ee%
Now, $||$ is the covariant derivative compatible with the
Christoffel and Ricci rotation coefficients.  $ T_{{\rm R}a}^{~~
\mu}=e_{a}^{\nu}T_{{\rm R}\ \nu}^{~\mu}$  the tetrad form of the
stress-energy tensor of Riemann curvature ${\cal
R}_{ab}^{~~\mu\nu}\in \mathfrak{so}(1,3)$.  It can be shown
\cite{wzc} that
\be\nno%
T_{{\rm R} \mu}^{~~\nu}
&=&{\cal R}_{ab \mu\la}{\cal R}^{ab\nu\la} - \frac{1}{4}\dl_\mu^\nu({\cal R}_{ab\la\ka}{\cal R}^{ab\la\ka})\nno \\
&=& \d 1 2 ({\cal R}_{\ka\si \mu\la}{\cal R}^{\ka\si \nu\la}+{\cal R}^*_{~\ka\si \mu\la} {\cal R}^{*\ka\si \nu\la}) \nno \\
&=&2C_{\la\mu}^{~~\ka\nu}{\cal R}^\la_\ka +\frac{{\cal R}}{3}({\cal R}_\mu^\nu -\frac{1}{4}{\cal R}\dl_\mu^\nu), \label{emR}
\ee%
where ${\cal R}_{\ka\si\mu\la}$ is the Riemann curvature tensor,
${\cal R}^*_{\ka\si\mu\la} =\frac 1 2 {\cal R}_{\ka\si \tau
\rho}\eps^{\tau \rho}_{\quad \mu\la}$ is the right dual of the
Riemann curvature tensor, $C_{\la\mu\ka\nu}$ is the Weyl tensor.  In
the last step in (\ref{emR}), the G\'eh\'eniau-Debever decomposition
for the Riemann curvature,
\be
{\cal R}_{\mu\nu\ka\la}=C_{\mu\nu\ka\la} + E_{\mu\nu\ka\la} + G_{\mu\nu\ka\la},
\ee
is used {\cite{GD}}, where
\be
E_{\mu\nu\ka\la} &=& \d 1 2 (g_{\mu\ka} S_{\nu\la}+g_{\nu\la} S_{\mu\ka}-g_{\mu\la} S_{\nu\ka}-g_{\nu\ka} S_{\mu\la}), \qquad\\
G_{\mu\nu\ka\la} &=& \d {\cal R} {12}(g_{\mu\ka}g_{\nu\la}-g_{\mu\la}g_{\nu\ka}), \\
S_{\mu\nu} &=& {\cal R}_{\mu\nu}-\d 1 4 {\cal R}g_{\mu\nu}.
\ee

It is easy to see that for the \dS-space the `energy-momentum'-like
tensor in (\ref{emR}) vanishes and the \dS-space satisfies the
Einstein-Yang equation (\ref{Geq3}) and (\ref{Geq3'}). Therefore,
the 4-d Riemann-sphere is just the gravitational instanton in this
model. In fact, it can be proved that any solution of the ordinary
vacuum Einstein equation (with $\Lambda$ term) has vanishing
`energy-momentum'-like tensor and does exactly satisfy the
Einstein-Yang equation (\ref{Geq3}) and (\ref{Geq3'}). So, this
simple model can pass the observation tests on solar scale. It can
also be shown \cite{hg,Han} that Einstein-Yang equations admit a
`Big Bang' solution. Different from \GR, $T_{{\rm R} jk}$ could play
a role as `dark stuff'.

Returning to the field equations (\ref{Geq2}) and (\ref{Geq2'}), all
the terms other than the Einstein tensor $\mathcal{R}_a^{\mu}-\frac
1 2 \mathcal{R}e_a^\mu$ that can be picked up from
$F_{~a}^{\mu}-\frac 1 2 Fe_a^\mu$ by means of the relation between
the Lorentz connection $B^{ab}_{~~\mu}$ and the Ricci rotation
coefficient $\gamma^{ab}_{~~\mu}$ should play an important role as
some `dark stuff' in the viewpoint of \GR. Thus, the model may
provide a more reasonable framework for the analysis of dark stuff
in the `precise cosmology'. Furthermore, since the field equations
are of the Yang-Mills-type, there should be gravitational potential
waves of both metric and Cartan connection, including the
gravitational metric waves in \GR.

In fact, this simple model can be viewed as a kind of \dS-gravity
in a `special gauge' and the 4-d pseudo-Riemann-Cartan manifolds
should be a kind of 4-d umbilical manifolds with local \dS-space
together with `gauged' \dS-algebra anywhere and anytime. In this
model, there is the $\La$ from the `gauge' symmetry so that it is
not just a `dummy' constant at classical level as in \GR. It is
important that even in this simple model the \dS-gravity is
characterized by a dimensionless coupling constant $g$ with
$g^2\sim G\hbar c^{-3}R^{-2}$. This supports the Planck
scale-$\Lambda$ duality.

Note that with the help of the connection (\ref{dSc}) valued in
$\mathfrak{so}(1,4)$, different gravitational dynamics can be
constructed. For example, MacDowell and Mansouri's approach %in
\cite{MM}, the dynamics is different from here. Recently, via a
symmetry broken mechanics for the the connection (\ref{dSc}) from
the \dS-algebra down to the Lorentz algebra has been explored
\cite{Lec} following \cite{SW}, and via a BF-type topological theory
Einstein's equation with cosmological constant in general relativity
has been discussed (see, e.g. \cite{FS}).

%%%%%%%%%%%%%%%%%%%%%%%%%%%%%%%%%%%%%%%%%%%%%%%

\section{Concluding  Remarks}

With plenty of \dS-puzzles, the dark universe as an accelerated
expanding one, asymptotic to a de Sitter-space with a tiny
cosmological constant $\Lambda$ \cite{Riess98,WMAP} greatly
challenges Einstein's theory of relativity as the foundation of
physics in large scale. It is well known that symmetry and its
localization play extreme important roles in modern physics. It
should also be the case in large scale.

It is important that there should be three kinds of special
relativity \cite{Lu,LZG,BdS,BdS05,IWR,T,NH,Lu05,PRdual,PoI,C3, Yan}
and their contractions \cite{NH}, and correspondingly there should
be also three kinds of theory of gravity as localization of the
relevant special relativity with some gauge-like dynamics,
respectively. While, the Nature may pick out one of them.

We have shown that there is a one-to-one correspondence between
Snyder's quantized space-time model and \dS\ special relativity. In
addition, a minimum uncertainty-like argument indicated by a
`tachyon' dynamics. Based on this correspondence and the argument,
we have made a conjecture that there should be a duality in physics
at the Planck scale and the cosmological constant, and there is
in-between the gravity characterized by a dimensionless constant
$g$.

Gravity in-between these two scales should be based on the
localization of \dS\, special relativity with a gauge-like dynamics
of full localized symmetry characterized by the dimensionless
coupling constant. A simple model of \dS-gravity in the Lorentz
gauge on umbilical manifold of local \dS-invariance may support this
point of view.

As the asymptotic behavior of our universe is no longer flat,
rather quite possibly a Robertson-Walker-like \dS-spacetime, our
universe may already indicate that \dS\, special relativity and
\dS-gravity with local \dS-invariance should be the foundation of
physics in the large scale.

{Finally, it should be stressed that the cosmological constant
$\Lambda=3R^{-2}$ is regarded as a fundamental constant and chosen
as the dark energy in cosmological observation. The latter is just a
working hypothesis. In fact, it is unnecessary to assume that the
entire dark energy is composed of the cosmological constant because
our universe should be described based on the gravity with local
\dS-invariance rather than \dSR\ itself.}

%\vskip 2mm \no
\begin{acknowledgments}

We'd like to thank Professors Z. Chang, Q.K. Lu, X.C. Song, S.K.
Wang, K. Wu and M.L. Yan, and Dr. X.N. Wu for valuable discussions.
This work is partly supported by NSFC under Grant Nos. 90403023,
90503002, 10505004, 10547002, 10605005.
\end{acknowledgments}

\appendix

{\section{Three Kinds of Beltrami Coordinate Systems}}
\label{sect:3-Beltrami}

First let us take the 2-d sphere as an example to illustrate the
intuitive geometric picture of what the Beltrami coordinates mean.
We often use a method called gnomonic projection to draw a map,
i.e., the projection of the earth's surface from the earth's center
to any plane not passing through the center. Under this kind of
projection, the great circles on the earth turn into straight lines
on the plane. Now if we consider the pseudo-sphere \HlsM\
(\ref{qiu2}), the hyperplane of projection has three cases:
timelike, spacelike and null\footnote{Hereafter the terms timelike,
spacelike and null for the hyperplane means that the signature of
the hyperplane is non-definite, {negative} definite and degenerate,
respectively.}. (With a slight abuse of terminology we also call the
corresponding (pseudo-)gnomonic projection to be timelike, spacelike
and null, respectively.) It can be proved, anyway, that the
geodesics on the \dS\ spacetime $H_l$ turn into straight (world)
lines on the hyperplane under this kind of (pseudo-)gnomonic
projection. These three kinds of hyperplanes of projection
correspond to three kinds of Beltrami coordinate systems, which we
will describe below.

\begin{figure}[!hbt]
\begin{center}
  \includegraphics{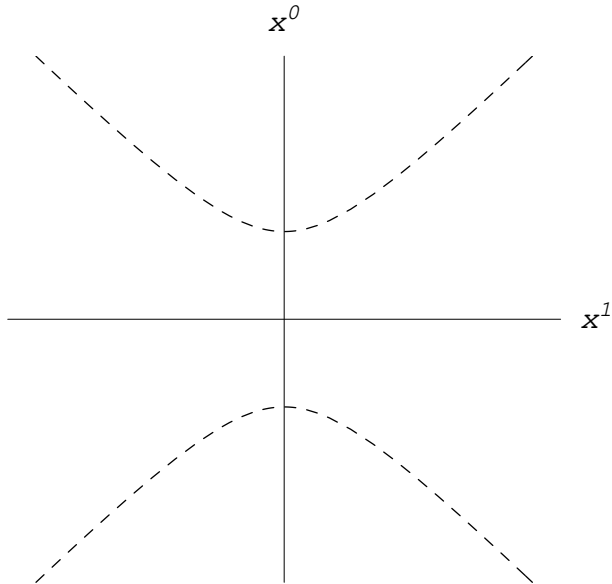}
  \caption{The 2-d illustration of the Beltrami coordinates from the timelike gnomonic projection,
  where the hyperbola $\sigma(x)=0$ is nearly one half of the projective boundary of the \dS\ spacetime, and the interior $\sigma(x)>0$ contains the points in the \dS\ spacetime.}
  \label{fig:Beltrami-TL}
\end{center}
\end{figure}

In the case of a timelike hyperplane of projection, since any
timelike hyperplane can be  transformed to the hyperplane
$\xi^4={const}$ under an $SO(1,4)$ transformation, we can only
consider the `gnomonic' projection to the hyperplane $\xi^4=l$,
without loss of generality. If the antipodal identification has not
been made, the `gnomonic' projection is, basically, {two-to-one}:
the $\xi^4>0$ part and $\xi^4<0$ part in the \dS\ spacetime $H_l$
have the same image on the hyperplane $\xi^4=l$. In order to make a
coordinate patch, we first consider the $\xi^4>0$ part (called the
$U^+_4$ patch), and then increase the number of patches to cover the
rest {portion}. Let $P^+_4$ denote the hyperplane $\xi^4=l$. Using
the mathematical language, every point $\xi\in H_l$ with $\xi^4>0$
has a one-to-one corresponding point $x\in P^+_4$, whose coordinates
are
\begin{equation}\label{Beltrami}
x^\mu\equiv\xi_x^\mu = l\frac{\xi^\mu}{\xi^4}, \quad \xi_x^4 = l.
\end{equation}
These coordinates $x^\mu$ are just the inverse Wick rotated version
of the Beltrami coordinates (\ref{Bcrd}). From the definition
(\ref{qiu2}) of $H_l$, it is easy to show that $x^\mu$ must satisfy
the condition (\ref{domain}) and the metric (\ref{metric}) follows.
On the hyperboloid $\sigma(x)=0$ of two sheets, the metric becomes
singular. In fact, $\sigma(x) = 0$ is nearly one half of the
projective boundary of $H_l$ \cite{GHTXZ} (see Figure
\ref{fig:Beltrami-TL}).\footnote{The other half of the
projective boundary of $H_l$ can be nearly described by
$\sigma(x)=0$ with the corresponding Beltrami coordinates on the
$U_4^-$ patch (see below).} The inverse of the metric (\ref{metric})
is
\begin{equation}
  g^{\mu\nu} = \sigma(x)(\eta^{\mu\nu} - l^{-2} x^\mu x^\nu).
\end{equation}
Now we consider the $\xi^4<0$ part (called the $U^-_4$ patch). It is
obvious that this part can be gnomonically projected to the
hyperplane $\xi^4=-l$ (called $P^-_4$), with the corresponding
Beltrami coordinates
\begin{equation}
  x^\mu \equiv -l\frac{\xi^\mu}{\xi^4}.
\end{equation}
It is not difficult to imagine that there should be at least eight
such patches $\{U_\alpha^\pm,\alpha=1,\cdots,4\}$ to cover the whole
$H_l$, with the Beltrami coordinates for $\alpha=1,2,3$ defined as
\begin{equation}
  x^\nu \equiv \pm l\frac{\xi^\nu}{\xi^\alpha},\quad
  \nu=0,\cdots,\hat{\alpha},\cdots,4,\quad \xi^\alpha\gtrless 0.
\end{equation}
{The hyperplanes of projection corresponding to $U^\pm_\alpha$
$(\alpha = 1, \ldots, 4)$ are timelike.} Evidently, {these} patches
can be related by $SO(1,4)$ transformations.

The Beltrami coordinate systems have intimate relation with
projective geometry. In fact, the Beltrami metric (\ref{metric}) can
be obtained purely through methods in projective geometry, where the
projective boundary (the hyperboloid $\sigma(x)=0$ of two sheets for
the case of timelike `gnomonic' projection above) is just the
so-called absolute \cite{GHTXZ}. The methods in projective geometry
clearly shows that, under the Beltrami coordinates, timelike, null
or spacelike geodesics in the \dS\ spacetime are straight lines
crossing, tangent to or apart from the absolute, respectively.

This conclusion holds not only for the case of timelike `gnomonic'
projection, but also for the case of null and spacelike `gnomonic'
projections that will be discussed below, the only difference being
the distinct absolute in each case. The Beltrami coordinates from
the timelike `gnomonic' projection has the advantage over the ones
from the null and spacelike `gnomonic' projections that all the
relevant expressions tend to their \Mink-counterparts under the flat
limit $l\to\infty$.

\begin{figure}[!hbt]
\begin{center}
  \includegraphics{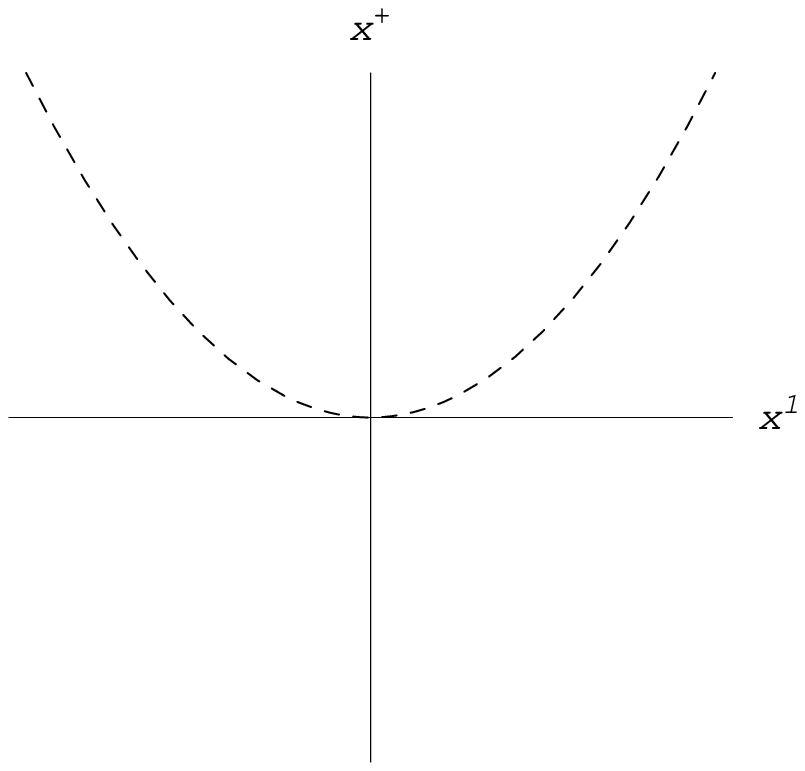}
  \caption{The 2-d illustration of the Beltrami coordinates from the null
  gnomonic projection, where the parabola $\varsigma(x)=0$ is  nearly one sheet of
  the projective boundary of the \dS\ spacetime, and the lower region
  $\varsigma(x)<0$ contains the points in the \dS\ spacetime.}
  \label{fig:Beltrami-LL}
\end{center}
\end{figure}

In the case of a null hyperplane of projection, it is convenient to
defined the following `light-cone' coordinates:
\begin{equation}
\xi^+\equiv\frac{\xi^0+\xi^4}{\sqrt{2}},\quad\xi^-\equiv\frac{\xi^0-\xi^4}{\sqrt{2}}.
\end{equation}
As well, without loss of generality, we can only consider the
`gnomonic' projection to the hyperplane $\xi^-=l$. The corresponding
Beltrami coordinates on the $\xi^->0$ patch are
\begin{equation}
  x^\mu \equiv l\frac{\xi^\mu}{\xi^-}, \quad \mu=+,1,2,3,
\end{equation}
where $x^\mu$ satisfy\footnote{In this paper, we use bold italic
letters $\vect{x},\vect{y}$ etc to stand for 3-d vectors.}
\begin{equation}
\varsigma(x) \equiv 2l x^+ - \vect{x}^2 < 0.
\end{equation}
Under this kind of Beltrami coordinates, the metric reads
\begin{equation}
ds^2=l^2\varsigma^{-1}(x)d\vect{x}^2+l^2\varsigma^{-2}(x)(l
dx^+-\vect{x}\cdot d\vect{x})^2.
\end{equation}
On the paraboloid $\varsigma(x) = 0$, which is nearly one
sheet\footnote{The projective boundary of the \dS\ spacetime
has topology $S^3\times\mathbb{Z}_2$.} of the projective boundary of
the \dS\ spacetime (see Figure \ref{fig:Beltrami-LL}), the metric
becomes singular. The corresponding inverse metric is
\begin{equation}
(g^{\mu\nu})=l^{-3}\varsigma(x)\left(%
\begin{array}{cc}
  2x^+ & x^i \\
  x^j & l\delta^{ij} \\
\end{array}%
\right).
\end{equation}

\begin{figure}[!hbt]
\begin{center}
  \includegraphics{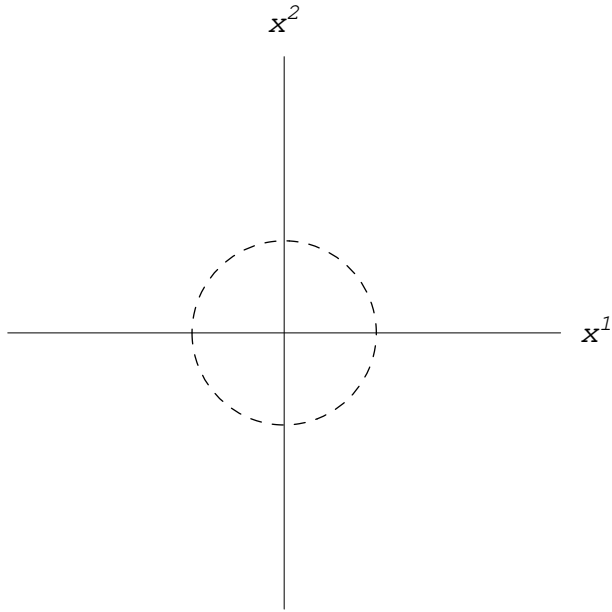}
  \caption{The 2-d illustration of the Beltrami coordinates from the spacelike gnomonic projection,
  where the circle $\hat\sigma(x)=0$ is one sheet of the projective boundary of the \dS\ spacetime, and the exterior $\hat\sigma(x)<0$ contains the points in the \dS\ spacetime.}
  \label{fig:Beltrami-SL}
\end{center}
\end{figure}

In the case of a spacelike hyperplane of projection, again we can
only consider the `gnomonic' projection to the hyperplane $\xi^0=l$,
with the corresponding Beltrami coordinates on the $\xi^0>0$ patch:
\begin{equation}
  x^\alpha \equiv l\frac{\xi^\alpha}{\xi^0}, \quad \alpha=1,2,3,4,
\end{equation}
where $x^\alpha$ satisfy
\begin{equation}
  \hat{\sigma}(x) \equiv 1 - l^{-2}\delta_{\alpha\beta}x^\alpha x^\beta<0.
\end{equation}
Under this kind of Beltrami coordinates, the metric is
\begin{equation}
ds^2=[\delta_{\alpha\beta}\hat{\sigma}^{-1}(x)+l^{-2}\delta_{\alpha\gamma}\delta_{\beta\delta}x^\gamma
x^\delta\hat{\sigma}^{-2}(x)]dx^\alpha dx^\beta.
\end{equation}
On the 3-sphere $\hat\sigma(x)=0$ the metric becomes singular,
which is one sheet of the projective
boundary of the \dS\ spacetime (see Figure \ref{fig:Beltrami-SL}).
The corresponding inverse metric is
\begin{equation}
  g^{\alpha\beta} = \hat{\sigma}(x)(\delta^{\alpha\beta} - l^{-2} x^\alpha x^\beta).
\end{equation}

%%%%%%%%%%%%%%%%%%%%%%%%%%%%%%%%%%%%%%%%%%%%%%%%%%%%%%%%%%%%
%%%% Fock's Theorem                         %%%%
%%%% Copied from BdS/BdS_axiom/rel_axiom.0.3.tex .  %%%%
%%%%%%%%%%%%%%%%%%%%%%%%%%%%%%%%%%%%%%%%%%%%%%%%%%%%%%%%%%%%
\section{On the General Transformations among Inertial Motions}
\label{sect:FockThm}

%\begin{thm}
{\bf Theorem (Fock's theorem):}

 If, under a coordinate transformation
$(x^i) \rightarrow (x'^i)$ on an $n$-dimensional region, a straight
line $x^i(\lambda) = x^i_0 + \lambda v^i$ (where $v^i$'s are
constants) is always transformed to be another straight line, then
there must be constants $A^i_{\ j}$, $B^i$, $C_i$ and $C$ such that
\begin{equation}
  x'^i = \frac{A^i_{\ j} x^j + B^i}{C_k x^k + C}
\label{eq:fraclinear}
\end{equation}
and
\begin{equation}
  \left| \begin{array}{rrcrr}
    A^1_{\ 1} & A^1_{\ 2} & \ldots & A^1_{\ n} & B^1 \\
    A^2_{\ 1} & A^2_{\ 2} & \ldots & A^2_{\ n} & B^2 \\
    \vdots & \vdots & \ddots & \vdots & \vdots \\
    A^n_{\ 1} & A^n_{\ 2} & \ldots & A^n_{\ n} & B^n \\
    C_1 & C_2 & \ldots & C_n & C
  \end{array} \right|
  \neq 0.
\label{eq:det}
\end{equation}
%\end{thm}

\begin{proof}
As assumed, there should be constants $v'^i$ and $\lambda' =
\lambda'(\lambda)$ such that the straight line $x^i(\lambda) = x^i_0
+ \lambda v^i$ are transformed to be a straight line $x'^i(\lambda')
= x'^i_0 + \lambda' v'^i$, namely,
\[
  x'^i(x_0 + \lambda v) = x'^i(x_0) + \lambda'(\lambda) v'^i.
\]
Differentiating it twice with respect to $\lambda$, we obtain \be &&
v^j \, \frac{\partial x'^i}{\partial x^j}(x_0 + \lambda v)
  = v'^i \, \frac{d\lambda'}{d\lambda}, \nno \\
&&  v^j v^k \, \frac{\partial^2 x'^i}{\partial x^j \partial x^k}
  (x_0 + \lambda v) = v'^i\,\frac{d^2 \lambda'}{d\lambda^2}. \nno
\ee Since we can always assume $d\lambda'/d\lambda > 0$, the above
equations give
\[
  v^j v^k \, \frac{\partial^2 x'^i}{\partial x^j \partial x^k} (x_0 + \lambda v)
  = v^j\,\frac{\partial x'^i}{\partial x^j} \,
    \frac{\,\frac{d^2 \lambda'}{d\lambda^2}\,}{\frac{d\lambda'}{d\lambda}}.
\]
For any point $(x^i)$, there is always a straight line passing
through it, with the tangent vector $v^i \frac{\partial}{\partial
x^i}$ at $(x^i)$. Obviously, $\frac{d^2
\lambda'}{d\lambda^2}/\frac{d\lambda'}{d\lambda}$ at $(x^i)$ depends
not only on $(x^i)$ but also on $v^i$.  Therefore, there will be a
function $f(x,v)$ such that
\[
  v^j v^k \, \frac{\partial^2 x'^i}{\partial x^j \partial x^k}
  = v^j\,\frac{\partial x'^i}{\partial x^j} \,f(x,v).
\]
Observe the above equation. We see that $f(x,v)$ must be linear for
$v^i$. Thus, there are functions $f_i(x)$ such that
\begin{equation}
  \frac{\partial^2 x'^i}{\partial x^j \partial x^k}
  = \frac{\partial x'^i}{\partial x^j}\,f_k(x)
  + \frac{\partial x'^i}{\partial x^k}\,f_j(x).
\label{eq:2der}
\end{equation}

The Jacobian $J(x) = \det\big(\frac{\partial x'^i}{\partial
x^j}\big)$ is nonzero everywhere. Its partial derivatives can be
obtained in the standard way as
\[
  \frac{\partial J}{\partial x^i}
  = \frac{\partial^2 x'^j}{\partial x^k \partial x^i}\,
    \Delta^k_j
  = \frac{\partial^2 x'^j}{\partial x^k \partial x^i}\,
    \frac{\partial x^k}{\partial x'^j} \, J,
\]
where $\Delta^k_j = \frac{\partial x^k}{\partial x'^j}\,J$ is the
algebraic complement of $\frac{\partial x'^j}{\partial x^k}$ in $J$.
Applying Eq.~(\ref{eq:2der}) to the above yields
\begin{equation}
  f_k = \frac{1}{n+1} \frac{\partial}{\partial x^k}\ln|J|.
\label{eq:f_mu}
\end{equation}
%where $n$ is the dimension of the manifold.

{From} Eq.~(\ref{eq:2der}) we obtain
\begin{widetext}
\[
  \frac{\partial^3 x'^i}{\partial x^j \partial x^k \partial x^l}
= \frac{\partial x'^i}{\partial x^j} \frac{\partial f_k}{\partial
x^l}
  + \frac{\partial x'^i}{\partial x^k} \frac{\partial f_j}{\partial x^l}
  + \frac{\partial x'^i}{\partial x^j} f_k f_l
  + \frac{\partial x'^i}{\partial x^k} f_j f_l
  + 2\,\frac{\partial x'^i}{\partial x^l} f_j f_k.
\]
Cycle the indices $j$, $k$ and $l$ in the above and add them
together. We thus have an expression of \be
  \frac{\partial^3 x'^i}{\partial x^j \partial x^k \partial x^l}
  & = & \frac{1}{3}\left[\frac{\partial x'^i}{\partial x^j}
    \left(\frac{\partial f_l}{\partial x^k}
    + \frac{\partial f_k}{\partial x^l} \right)
+ \frac{\partial x'^i}{\partial x^k}
    \left(\frac{\partial f_j}{\partial x^l}
  + \frac{\partial f_l}{\partial x^j} \right)
  + \frac{\partial x'^i}{\partial x^l}
    \left(\frac{\partial f_k}{\partial x^j}
    + \frac{\partial f_j}{\partial x^k}\right)
\right. \nno
\\
  & & \left. + 4 \,\frac{\partial x'^i}{\partial x^j} \, f_k f_l
  + 4 \,\frac{\partial x'^i}{\partial x^k} \, f_l f_j
  + 4 \,\frac{\partial x'^i}{\partial x^l} \, f_j f_k
  \right].
\ee
\end{widetext}
Using the above two equations we obtain
\[
  \frac{\partial f_k}{\partial x^j}
  + \frac{\partial f_j}{\partial x^k}
  = 2\,f_j f_k.
\]
Since the two terms on the left hand side are equal (see,
Eq.~(\ref{eq:f_mu})), we have
\begin{equation}
  \frac{\partial f_k}{\partial x^j} = f_j f_k.
\label{eq:f_sigma_rho}
\end{equation}

Using the above equation and Eq.~(\ref{eq:f_mu}), we can verify that
the partial derivatives of $f_i\,|J|^{-1/(n+1)}$ are zero. Thus
there are constants $C_i$ such that
\begin{equation}
  f_i = - C_i\,|J|^{1/(n+1)}.
\end{equation}
Substituting it into Eq.~(\ref{eq:f_mu}) we can solve
\begin{equation}
  |J|^{-1/(n+1)} = C_i x^i + C
\end{equation}
with $C$ an integral constant. And, from eqs.~(\ref{eq:2der}),
(\ref{eq:f_mu}) and (\ref{eq:f_sigma_rho}) we can check that
\begin{equation}
  \frac{\partial^2}{\partial x^j \partial x^k}
  \Big( x'^i |J|^{-1/(n+1)}\Big) = 0.
\end{equation}
Therefore, Eq.~(\ref{eq:fraclinear}) is satisfied.
%In order that $J \neq 0$, (\ref{eq:det}) must be satisfied.

It can be verified that, if the coordinate transformation
Eq.~(\ref{eq:fraclinear}) is %with (\ref{eq:det}) being
satisfied, then a straight line is always transformed to be a
straight line.
\end{proof}

The Jacobian of the transformation (\ref{eq:fraclinear}) is, in
fact,
\begin{equation}
  J(x) = \frac{D}{(C_i x^i + C)^{n+1}},
\qquad
  D = \left|
  \begin{array}{ll}
    A^i_{\ j} & B^i \\
    C_j & C
  \end{array}
  \right|.
\end{equation}
In the proof, we have obtain that $|J| = 1/(C_i x^i + C)^{n+1}$.
This implies that $D = \pm 1$ in the proof. In fact, we can always
require that
\begin{equation}
  D = 1
\end{equation}
in the transformation (\ref{eq:fraclinear}).

\end{document}